\documentstyle[aps,prl,epsfig,epic,eepic,array,floats]{revtex}
\draft

\begin{document}

\def\Real{\mathop{\rm Re}\nolimits}
\def\bvec#1{{\bf #1}}
\def\vhat#1{\hat{\bvec{#1}}}

\title{Influence of diffraction on the spectrum and wavefunctions of
  an open system}

\author{ J. S. Hersch$^*$, M. R.  Haggerty$^*$, and E. J.
  Heller$^\dagger$ \\}

\address{
  $^*$Department of Physics, Harvard University, Cambridge,
  Massachusetts 02138\\
  $^\dagger$Departments of Chemistry and Physics, Harvard University,
  Cambridge, Massachusetts 02138}

\maketitle
\begin{abstract}
    In this paper, we demonstrate the existence and significance of
    diffractive orbits in an open microwave billiard, both
    experimentally and theoretically.  Orbits that diffract off of a
    sharp edge of the system are found to have a strong influence on
    the transmission spectrum of the system, especially in the regime
    where there are no stable classical orbits.  On resonance, the
    wavefunctions are influenced by both classical and diffractive
    orbits.  Off resonance, the wavefunctions are determined by the
    constructive interference of multiple transient, nonperiodic
    orbits.  Experimental, numerical, and semiclassical results are
    presented.
\end{abstract}

\section{Introduction}

In this work, we discuss the transmission spectrum and wavefunctions
of an open resonator coupled to a quantum point contact (QPC).  The
system exhibits both stable and unstable dynamics, depending on the
value of a single parameter. The spectral properties of the resonator
are determined by the interference of closed (not necessarily
periodic) orbits that begin and end at the QPC.  Semiclassically, one
computes the transmission of such a system with a sum over the {\em
  classical} trajectories.  However, it was found experimentally that
there were many resonances in the spectrum which did not appear in the
theory when classical trajectories involving only specular reflection
were considered.  We found that the missing resonances were only
reproduced when {\em nonclassical} closed orbits that undergo {\em
  diffraction} were included into the semiclassical sum for the
transmission.

This issue of including diffraction into the semiclassical propagator
has been considered by various authors, for both closed
systems~\cite{vattay94,pavloff95,sieber97} and open
systems~\cite{whelan95}.  The basis for many of these treatments is
the geometric theory of diffraction, originated by
Keller~\cite{keller62}.  A distinguishing feature of the work
presented here is that, in the unstable regime of our resonator, the
effect of the diffractive orbits is of the same order as the purely
classical orbits.  A consequence of this is that there are as many (or
more) resonances that are supported by diffractive orbits as are
supported by simple classical orbits.  This is related to the fact
that there is only one classical closed orbit in our system in the
unstable regime.  This is in stark contrast to the case of a closed,
unstable (chaotic) system.  Normally, for a closed system, diffraction
plays a minor role because of the overwhelming number of
nondiffractive periodic orbits present.  However, when the system is
open, the great majority of long periodic orbits might vanish, if they
are allowed to escape the system.  In such a case, where only a very
small number of classical periodic orbits are present, diffractive
orbits may gain in importance and considerably affect the spectrum of
the system.  This situation is realized by our resonator.

The paper is organized as follows: in Sec.~\ref{sec:resonator}, we
discuss the resonator being studied.  In Sec.~\ref{sec:experiment}, we
describe the experimental apparatus.  In Sec.~\ref{sec:results}, we
present the experimental results, which are comprised of measured
spectra and wavefunctions.  In Sec.~\ref{sec:geometric} we provide a
short introduction to the geometric theory of diffraction, which is
the theory that describes how diffractive rays contribute to the
semiclassical description of a quantum mechanical wavefunction.  In
Sec.~\ref{sec:energy} we show in detail how the geometric theory of
diffraction is incorporated into the semiclassical trace formula.
Here theoretical results are seen to be in excellent agreement with
measured data.  In Sec.~\ref{sec:time} the physics of the resonator in
the time domain is discussed, and again very good agreement between
theoretical and experimental observations is found.

In Sec.~\ref{sec:avoided} we discuss the behavior of the system as the
transition between stable and unstable dynamics is crossed, as well as
analogies of the open resonator with some well-known closed systems,
namely the lemon and stadium billiards.  Finally, in
Sec.~\ref{sec:imaging} we discuss the possibility of imaging a pure
state quantum mechanical wavefunction in a mesoscopic system with the
help of an atomic force microscope, with concluding remarks in
Sec.~\ref{sec:conclusion}.  A short paper discussing the experimental
results presented here has been previously
published~\cite{hersch99.1}.  See also~\cite{mythesis}.

\section{The resonator}
\label{sec:resonator}
\begin{figure}
    \centerline{\epsfig{figure=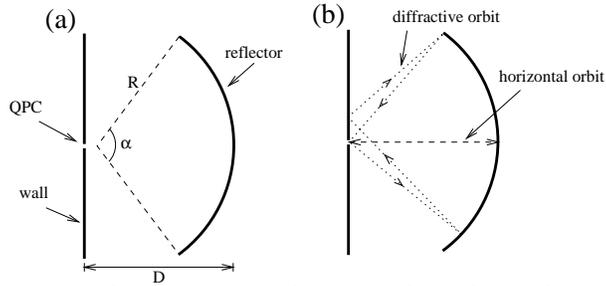, width=8cm}}
    \caption{
      (a) A schematic of the mesoscopic resonator studied by Katine,
      with radius of curvature $R$, opening angle $\alpha$, and
      reflector-wall separation $D$.  Electrons impinge on the wall
      from the left, and the conductance to the region on the right is
      measured.  (b) Two closed orbits of the unstable resonator:
      diffractive (dotted line), and horizontal (dashed line).  These
      will be discussed later in the paper.}
    \label{schematic}
\end{figure}

Recently, Katine studied the transmission behavior of an open quantum
billiard in the context of a two dimensional electron gas (2DEG) in a
GaAs/AlGaAs heterostructure~\cite{katine97}.  Their resonator was
formed by a wall with a small aperture (the QPC), and an arc-shaped
reflector.  A schematic of this resonator is shown in
Fig.~\ref{schematic}.  The voltage on the reflector could be varied,
effectively moving the reflector towards or away from the wall.  Their
measurements showed a series of conductance peaks, analogous to those
seen in a Fabry-Perot, as the reflector position was varied.

An interesting property of the resonator considered here is that it is
geometrically {\em open}, but in the stable regime it is classically
{\em closed}.  In the unstable regime, the resonance properties of the
billiard are determined in large part by {\em diffraction}.

The resonator shown in Fig.~\ref{schematic} has two distinct modes of
operation.  When the center of curvature of the reflector is to the
left of the wall (the regime studied in~\cite{katine97}), then almost
all classical paths starting from the QPC that hit the reflector
remain forever in the region between wall and the reflector: the
dynamics is stable and the periodic orbits can be semiclassically
quantized.  Each quantized mode of the resonator can be characterized
by two quantum numbers $(n,m)$, which represent the number of radial
and angular nodes respectively.  As the reflector-wall separation is
varied, the conductance exhibits a peak each time one of these
quantized modes is allowed.  Once an electron is in the resonator, the
only way for it to leave is by tunneling back through the QPC or by
diffracting around the reflector; since both processes are slow, the
resonances have narrow widths.  Because the QPC is on the symmetry
axis, only modes with even $m$ are excited significantly.  The states
of the resonator in the stable regime bear a strong resemblance to a
certain symmetry class of states in a lemon-shaped
billiard~\cite{postmodern}.  This class has even symmetry about the
short axis of the lemon, and odd symmetry about the long axis.  This
connection will be explored more fully in Section~\ref{sec:avoided}.

When the center of curvature is to the right of the wall, however, the
dynamics becomes unstable: all classical trajectories beginning at the
QPC rapidly bounce out of the resonator, except for a single unstable
periodic orbit along the axis of symmetry, which we will call the
``horizontal'' orbit [see Fig~\ref{schematic}(b)].  The horizontal
orbit is a member or a class of orbits that we call ``geometric,''
because their paths are governed by specular reflection off the wall
and reflector, and do not undergo diffraction.  Although the
horizontal orbit returns to the QPC, the electron has a low
probability of escaping the resonator there because the QPC is much
smaller than the de Broglie wavelength of the electron.  Because the
horizontal orbit is the only periodic orbit in the unstable regime,
one might expect resonant buildup only along the symmetry axis.  Such
a spectrum would be quasi-one-dimensional, with only the
half-wavelength periodicity of a Fabry-Perot cavity.  However, in
numerical simulations it was found that there are other transmission
resonances in the unstable regime which did not correspond to any
classical periodic orbits~\cite{edwardsthesis}.  It was proposed that
these anomalous peaks are supported by diffraction off the tips of the
reflector.  Unfortunately, in the mesoscopic experiments, decoherence
of the electron wave by impurities in the GaAs/AlGaAs heterostructure
shortens the lifetime of the resonances, leaving insufficient energy
resolution to resolve the diffractive peaks~\cite{westervelt_private}.

\section{Experiment}
\label{sec:experiment}
\begin{figure}
    \centerline{\epsfig{figure=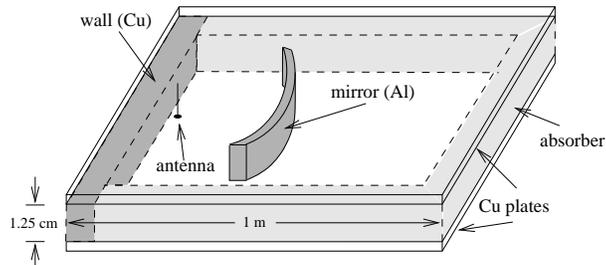, width=8cm}}
    \caption{
      The microwave analog of the mesoscopic resonator studied by
      Katine.  The antenna simulates the QPC; to reduce its coupling
      to the resonator, it is placed very close to the wall.  The
      drawing is not to scale.  }
    \label{experiment}
\end{figure}

Because of the problems of dissipation and decoherence in the
mesoscopic experiments, we decided to investigate a parallel plate
microwave resonator with a similar geometry.  In microwave
experiments, decoherence and dissipation are not a problem, the
geometry of the resonator can be specified much more accurately, and
the dynamical range of available wavelengths is much larger.  The
experimental setup is shown in Fig.~\ref{experiment}.
The equation governing the component of the electric field normal to
the plates for the TEM mode is identical to the two-dimensional
time-independent Schr\"odinger
equation~\cite{gokirmak98,sridhar92,stein92,stockman90}.  Therefore,
by studying the modes of parallel-plate resonators we can gain insight
into the behavior of two-dimensional solutions to the Schr\"odinger
equation.

The resonator consisted of two parallel copper plates, 1 meter square,
separated by a distance of 1.25~cm.  One side of the resonator
consisted of a copper wall.  The other three sides were lined with a
11.5~cm thick layer of microwave absorber (C-RAM LF-79, Cuming
Microwave Corp.)\ designed to provide 20~dB of attenuation in the
reflected wave intensity in the range 0.6-40~GHz.  The absorber
prevented outgoing waves from returning to the resonator, thereby
simulating an open system in the directions away from the wall.  An
antenna was inserted normal to the plates, 2~mm from the wall, to
simulate the QPC.  The curved reflector was formed from a rectangular
aluminum rod bent into an arc with radius of curvature $R = 30.5$~cm.  
Various opening angles $\alpha$ were used: $115^\circ$,
$112^\circ$, $109^\circ$, and $106^\circ$.

Instead of measuring the transmission of the resonator, we measured
the reflection back from the antenna; for this we used an HP8720D
network analyzer in ``reflection'' mode (the complex $S_{11}$ parameter
of the resonator was measured).  We inferred the transmission
probability $|T|^2$ via $|T|^2 = 1 - |R|^2$, where $R = S_{11}$ is the
measured reflection coefficient.  Because of the proximity of the
antenna to the wall, it was only weakly coupled to the resonator;
therefore, in the absence of the reflector, the transmission
coefficient was close to zero.  However, when the reflector was
present, the transmission experienced pronounced maxima at certain
frequencies.  

\section{Results}
\label{sec:results}
\begin{figure}
    \centerline{\epsfig{figure=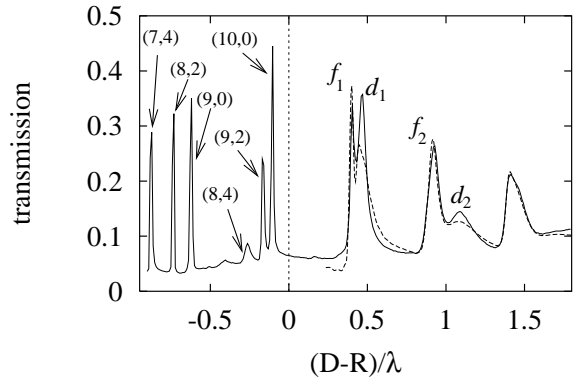,width=8cm}}
    \caption[Transmission vs. reflector-wall separation]{
      Experimental transmission versus reflector-wall separation at a
      fixed frequency of 5.63 GHz; i.e., $R = 5.7 \lambda$.  The
      stable/unstable transition point occurs at abscissa zero.  In
      the stable regime, the peaks are labeled by their radial and
      angular quantum numbers $(n,m)$.  In the unstable regime, the
      diffractive resonances (labeled $d$) appear to the right of the
      Fabry-Perot peaks (labeled $f$). The dashed curve is the result
      of a semiclassical calculation which takes diffractive orbits
      into account (see text). The opening angle for the reflector was
      $\alpha = 106^\circ$. With the reflector removed from the
      cavity, the transmission was 0.15 in these units.}
    \label{transmission}
\end{figure}

\begin{figure}
    \centerline{\epsfig{figure=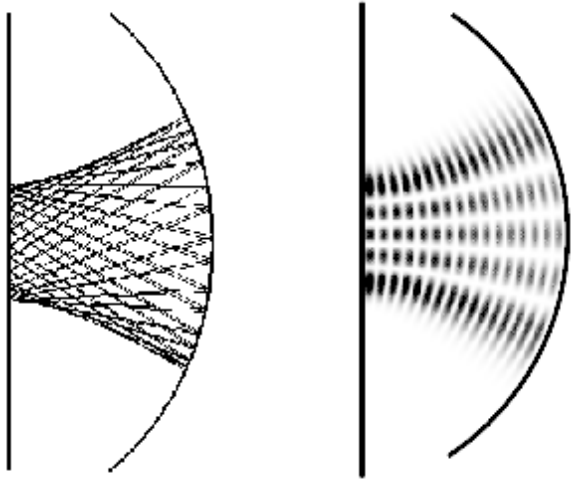,width=8cm}}
    \caption[A trajectory in the stable regime]
    {On the left is a classical trajectory starting at the QPC.  On
      the right is the corresponding computed wavefunction.}
  \label{fig:stablefig}
\end{figure}

In Fig.~\ref{transmission} we show a transmission spectrum at fixed
frequency, as the distance between the wall and reflector is varied.
In the stable regime, we see that the peaks are narrow and well
defined.  This is because the dynamics is stable in this regime:
nearly all trajectories starting from the QPC that hit the reflector
remain forever in the region between the wall and the reflector.  In
this regime, there exist invariant tori, which may be semiclassically
quantized to produce the states of the stable resonator.  Such a
classical orbit is shown in Fig.~\ref{fig:stablefig},
along with its quantum mechanical wavefunction counterpart.  We see
that the trajectory does not approach the region where the resonator
is open.  Thus, it behaves as if the cavity were {\em closed}---hence
the narrow widths of the peaks in the stable regime.

In the unstable regime, the transmission curve is quite different.
Here, there are two types of resonances.  The first type, labeled $f$
in Fig.~\ref{transmission}, is related to the horizontal orbit along
the axis of symmetry, and bears some resemblance to a Fabry-Perot type
resonance between two half-silvered mirrors.  The second type, labeled
$d$, is supported by diffraction off the tips of the reflector.

\begin{figure}
    \centerline{\epsfig{figure=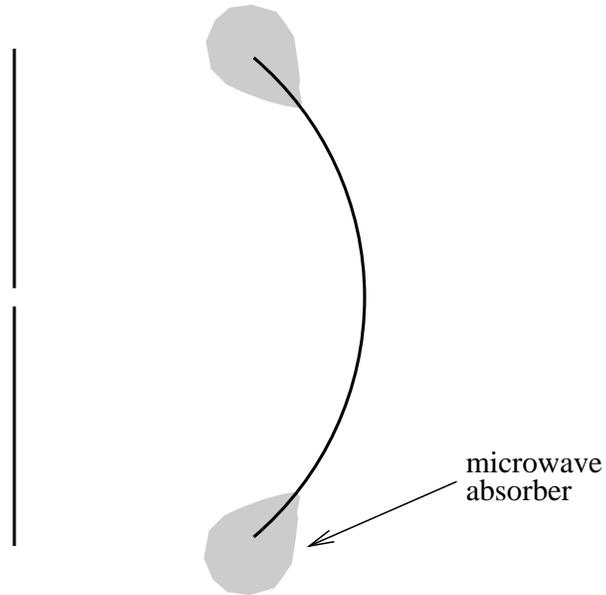,width=8cm}}
    \caption[Elimination of diffractive orbits by absorption]
    {Diffractive peaks could be removed from the spectrum by placing
      microwave absorber near the tips of the reflector, as
      indicated.}
    \label{fig:tips}
\end{figure}

We verified experimentally that the $d$-peaks were indeed supported by
diffraction by surrounding the tips of the reflector with microwave
absorber and repeating the experiment, as indicated in
Fig.~\ref{fig:tips}.  When this was done, the Fabry-Perot resonances
were unaffected, but the diffractive peaks were entirely eliminated
from the spectrum.  This makes sense because the gradually thickening
absorber smoothly attenuated reflections from rays coming near the
tip, leaving no sharp discontinuity from which rays could diffract.

The wavefunctions corresponding to peaks $f_1$ and $d_1$ were measured
using the technique of Maier and Slater~\cite{maier54}.  They showed
that the frequency shift of a given resonance due to a small sphere of
radius $r_0$ at a position $(x,y)$ is given by
\begin{equation}
    \frac{\omega^2 - \omega_0^2}{\omega_0^2} = 
    4\pi r_0^3 \left(\frac{1}{2}H_0^2(x,y) - E_0^2(x,y) \right), 
    \label{eq:sphere-shift}
\end{equation}
where $E_0$ and $H_0$ are the unperturbed electric and magnetic
fields.  Thus, the frequency shift is proportional to the local
intensity of the microwave field, and by measuring the shift as a
function of the position of the sphere, the field intensity of a
particular mode can be mapped out.  Note that the frequency shift is
positive in regions where the magnetic field is large, and negative
where the electric field is large.  Also, the factor of $1/2$
multiplying the magnetic field in Eq.~(\ref{eq:sphere-shift})
indicates that the sphere is a stronger perturbation to the electric
field than the magnetic field.  In our measurements, we found this to
be the case: the shifts were predominantly negative.  Appreciable
positive shifts were found only at the nodes of the electric field,
corresponding to maxima of the magnetic field.

\begin{figure}
    \centerline{\epsfig{figure=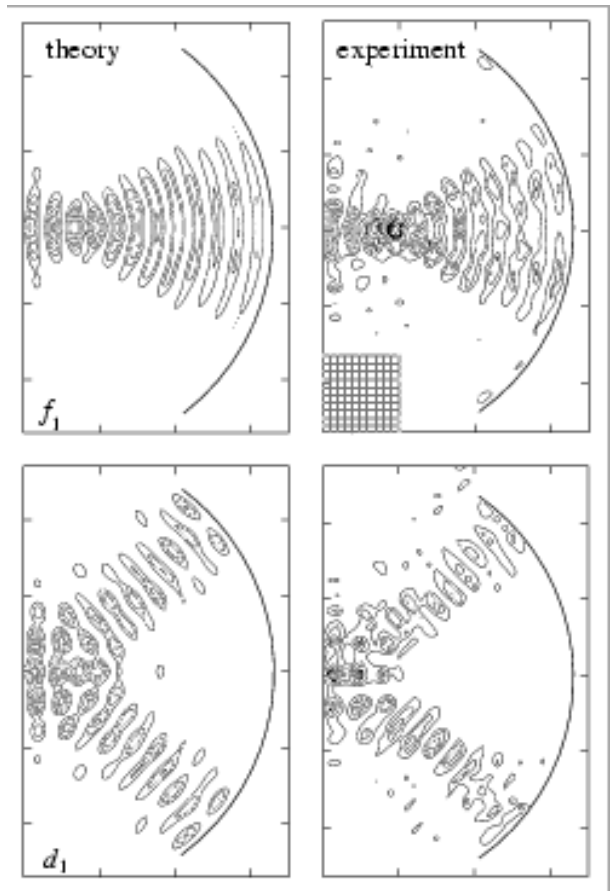,width=8cm}}
    \caption[Wavefunctions: theory and experiment]{
      Comparison between theoretical quantum wavefunctions (left) and
      experimentally measured microwave frequency shifts (right).  The
      two modes correspond to peaks $f_1$ and $d_1$, respectively, in
      Fig.~\ref{transmission}.  The wall is located on the left
      vertical axis in each plot, and the reflector position is
      indicated by the arc.  The graph ticks are 10 cm apart.  The
      fine grid indicates the spacing of the experimentally sampled
      points (grid spacing 1~cm).}
    \label{wavefunctions} 
\end{figure}

Figure~\ref{wavefunctions} shows theoretical quantum wavefunctions
compared with experimentally measured frequency shifts for the
resonances labeled by $f_1$ and $d_1$ in Fig.~\ref{transmission}.  The
theoretical wavefunctions were generated using Edwards' wavelet method
presented in~\cite{edwardsthesis}.
The measured frequency shift is plotted as a function of sphere
position.  For these measurements, we used a steel bead of diameter
4.0~mm for the perturbation.  The bead was rastered over the inside of
the cavity by means of an external magnet.  That way, the bead could
be moved around inside the cavity without taking the cavity apart.  It
is important to note that the frequency shift is not proportional to
$E^2$, but rather to $H^2/2 - E^2$.  Therefore we show only negative
contour lines below 20\% of the maximum negative shift, and thereby
emphasize regions of strong electric field.  The similarity between
theory and experiment is striking.

The wavefunction labeled $f_1$ in Fig.~\ref{wavefunctions} is clearly
associated with the horizontal orbit along the axis of symmetry.  Rays
emanating from a point source located on the axis of symmetry next to
the wall bounce off the reflector and come to an approximate focus
about 10~cm from the source.  The focus is approximate because of
spherical (or in this case cylindrical) aberration.

Now we turn our attention to the state labeled $d_1$ in
Fig.~\ref{wavefunctions}.  As noted above, the only periodic orbit in
the unstable regime is the horizontal orbit, along the axis of
symmetry.  The pictured wavefunction, however, clearly has very little
amplitude along this periodic orbit.  Instead the wavefunction has a
band of higher amplitude running from the region of the tip of the
mirror to the QPC, but in the unstable regime there is no {\em
  classical} periodic orbit that does this.  Later in the paper, it
will be shown that states such as $d_1$ are supported by orbits that
undergo {\em diffraction} off the tips of the reflector.  One such
orbit is shown in Fig.~\ref{schematic}(b).  Rays that hit the smooth
surfaces of the reflector or wall undergo specular reflection, whereas
the rays that hit near the reflector tips can be diffracted.  A
fraction of the wave amplitude can then return to the QPC from this
region, thus setting up a {\em non-classical\/} closed orbit.  All
peaks labeled with a $d$ in Fig.~\ref{transmission} are supported by
such diffractive orbits.  

Numerical calculations have shown that for energies off resonance, the
quantum wavefunction is often intermediate between those shown for
$f_1$ and $d_1$, in the sense that amplitude seems to be running from
the QPC to some point between the center and the tip of the
reflector~\cite{edwardsthesis}.

\begin{figure}
    \centerline{\epsfig{figure=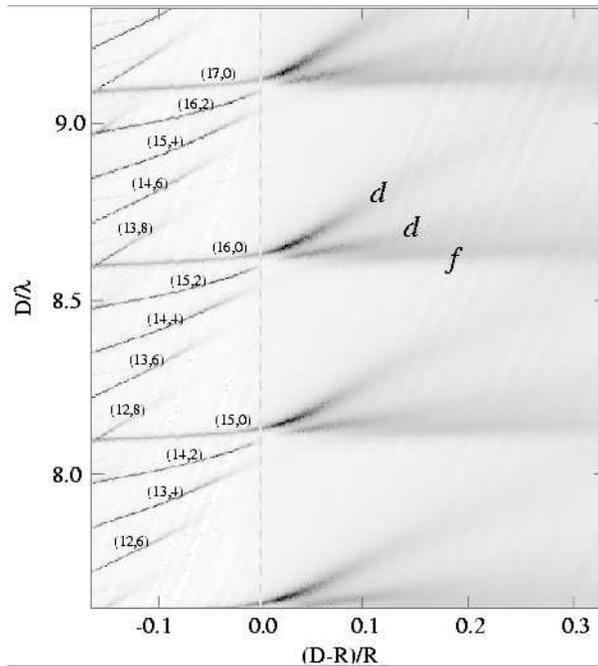,width=8cm}}
    \caption[Experimental transmission vs. wavelength and 
    reflector separation]
    {Experimental transmission versus reflector-wall separation and
      wavelength.  High transmission regions are dark.  On the left of
      the vertical dotted line is the stable regime, where the
      transmission peaks are sharp.  The quantum numbers $(n,m)$ are
      indicated for many of the peaks.  On the right is the unstable
      regime, where the resonances become wider and diffractive orbits
      become important.  Transmission peaks supported by diffractive
      orbits are marked by $d$.}
    \label{pretty}
\end{figure}

This can be understood in terms of the interference of paths with each
other as they ``walk off'' the horizontal orbit and escape the
resonator.  Thus diffraction does not necessarily play a major role in
determining the off-resonance wavefunctions.  However, diffraction
{\em is} instrumental in determining the on-resonance wavefunctions
underlying the conductance peaks $d_1$ and $d_2$ in
Fig.~\ref{transmission}.  Figure~\ref{pretty} shows a more global
picture of the transmission properties of the resonator.
Here we plot the transmission of the resonator as both the wavelength
and the reflector-wall separation are varied.  Each vertical slice
through this figure is a frequency spectrum with fixed reflector
position; the dotted line marks the classical transition from stable
to unstable motion that occurs when the reflector's center of curvature
moves to the right of the QPC.  The vertical axis indicates how many
wavelengths fit along the horizontal orbit between the QPC and the
reflector.  The repetition of the resonance pattern every
half-wavelength in the vertical direction is analogous to the
half-wavelength periodicity of a Fabry-Perot cavity.

In the stable regime the peaks have been labeled with their quantum
numbers, $(n,m)$.  Because of the choice of vertical axis, the $m = 0$
resonance peaks are approximately horizontal in this figure.  As the
stable/unstable transition is approached, the peaks with high $m$
disappear one by one because their large angular sizes allow them to
escape around the reflector.

At the stable/unstable transition, all of the resonances in a family
would be approximately degenerate, but instead there is an avoided
crossing.  The level repulsion is caused by a coupling that is partly
mediated by diffraction; this subject will be explored more thoroughly
in Section~\ref{sec:avoided}.

\begin{figure}
    \centerline{\epsfig{figure=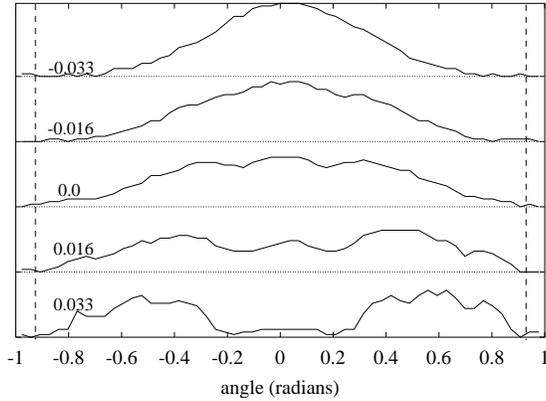,width=8cm}}
    \caption{In this figure, we show the angular dependence of a
      diffractive state as the reflector is moved through the
      stable/unstable transition.  Plotted is the amplitude of the
      frequency shifts measured 1 cm away from the reflector. The
      value of $(D-R)/R$ is indicated for each curve. The
      tips of the reflector are indicated on the plot by vertical
      dashed lines.}
    \label{fig:angular}
\end{figure}

In Fig.~\ref{fig:angular} the angular dependence of a diffractive
state is shown as the reflector is passed adiabatically through the
stable/unstable transition point. In the stable region, most of the
amplitude is along the symmetry axis.  As the reflector
moves further from the wall, the amplitude slowly separates into two
lobes, with very little in the center.  When the reflector is only
slightly in the unstable region, there are bands of amplitude running
to a point on the reflector intermediate between the center and the
tips, as in the curve for $(D-R)/R = 0.016$ in Fig.~\ref{fig:angular}.

In the unstable regime, the only remaining classical periodic orbit is
the horizontal orbit, which itself becomes unstable.  The Fabry-Perot
peak (labeled $f$) is essentially quantized along the horizontal
orbit, so its position shows a simple dependence on reflector
position.  It becomes broad in the unstable regime, with a lifetime
given by the classical Lyapunov stability exponent of the horizontal
orbit. Two diffractive resonances (labeled by $d$), are also visible
in each family; they separate from the Fabry-Perot type peak as the
reflector is moved away from the wall.

The diffractive peaks labeled by $d$ in Fig.~\ref{pretty} cannot be
explained by semiclassical theory unless diffraction off the tips of
the reflector is included, as will be shown in the following sections.

\section{Geometric theory of diffraction}
\label{sec:geometric}
Before we consider the problem of computing semiclassically the
transmission properties of our resonator, let us study the simpler
problem of diffraction of a plane wave off an infinite half-line in
2D.  This problem will serve as a good introduction into the
geometrical theory of diffraction, which will be used to include
diffraction into the semiclassical propagator.

\begin{figure}
    \begin{center}
        \centerline{\epsfig{figure=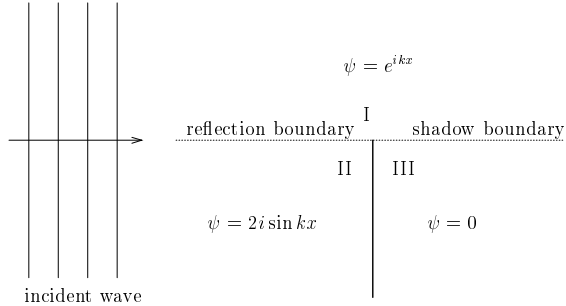,width=8cm}}
        \caption[Diffraction by a half-line]
        {Diffraction from an infinite half-line in 2D.}
        \label{fig:halfplane}
    \end{center}
\end{figure}

The problem is illustrated in Fig.~\ref{fig:halfplane}.  A plane wave
$e^{ikx}$ is normally incident on the half-line from the left. The
half-line extends up from the middle of the figure, indicated by the
dark line.  We take the tip of the line to be our coordinate origin.
Within the geometrical optics approximation, the problem is divided
into three separate regions: that of transmission, reflection, and
shadow, labelled I, II, and III, respectively.  The values of the
wavefunction in each region are indicated in the figure as well. In
region I, the wave does not hit the wall and thus is unchanged within
the geometric optics approximation.  In region II, the wave is
perfectly reflected and thus a standing sine wave is set up there.  In
region III, we have a perfect shadow region, which is completely dark.
Along the reflection and shadow boundaries indicated, the solution is
discontinuous. Of course, these discontinuities are not present in the
exact solution; they are an artifact of the geometric approximation.
As we shall see, it is the diffraction off the tip of the wall that
corrects these discontinuities.

\begin{figure}
    \begin{center}
        \centerline{\epsfig{figure=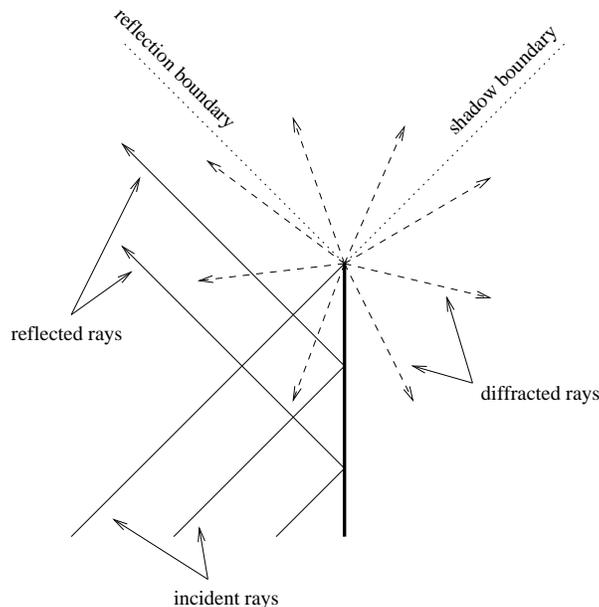,width=8cm}}
        \caption[Reflected and diffracted rays]{Reflected and
          diffracted rays from an infinite half-line.  Diffracted
          rays are shown in dashed lines.  The reflection and shadow
          boundaries are indicated. The wave is incident from the
          lower left corner of the figure, as indicated by the
          incident rays.}
        \label{fig:keller}
    \end{center}
\end{figure}

\begin{figure}
    \begin{center}
        \centerline{\epsfig{figure=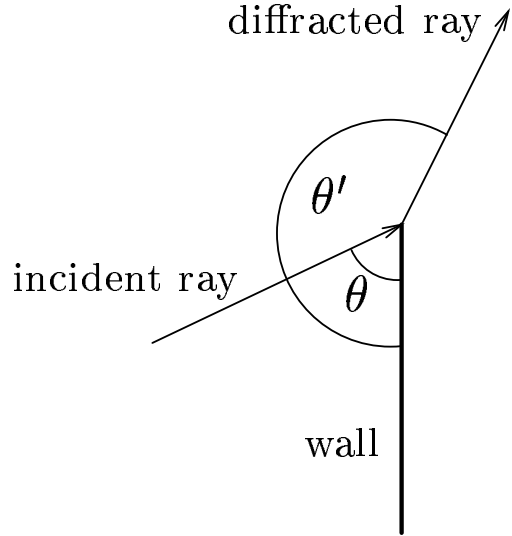,width=8cm}}
        \caption[Angles of incident and diffracted ray]
        {Angles of the incident and diffracted ray for use in
          Eq.~(\ref{eq:diff_coeff}). }
        \label{fig:angles}
    \end{center}
\end{figure}

In 1953, Keller showed that one can think of diffraction as
originating from a group of ``diffracted rays'' originating from the
edge of the wall~\cite{keller62}.  The idea is illustrated in
Fig.~\ref{fig:keller}.
Away from the shadow and reflection boundaries, these diffracted rays
have the form of an outgoing cylinder wave, multiplied by an angle
dependent ``diffraction coefficient.'' However, Keller's original
theory was shown to be invalid on the reflection and shadow
boundaries.  A properly uniformized geometric theory of diffraction
was developed by Kouyoumjian and Pathak~\cite{kouyoumjian74}.  The
diffracted rays are multiplied by a suitable complex number which
depends both on the angles of the incident and diffracted rays
relative to the wall, as well as the distance from the edge.  In this
uniformized theory, the solution to the half-line is given by
\begin{equation}
    \label{eq:uniform}
    \psi({\bf r}) = \psi_{\rm g}({\bf r}) +
    D(\theta,\theta',r,k)e^{ikr},
\end{equation}
where $\psi_{\rm g}({\bf r})$ is the solution given by geometrical
optics, shown in Fig.~\ref{fig:halfplane}. The diffraction coefficient
$D(\theta,\theta',r,k)$ is given by~\cite{james76,footnote1},
\begin{eqnarray}
    \label{eq:diff_coeff}
    D(\theta,\theta',r,k) &=& 
    -\mbox{sgn}(a_i)K\left(|a_i| \sqrt{kr}\right)
    +\mbox{sgn}(a_r)K\left(|a_r| \sqrt{kr}\right),\\
    a_{i,r} &=& \sqrt{2}\cos\left(\frac{\theta\mp\theta'}{2}\right),
\end{eqnarray}
and $K(x)$ is a modified Fresnel integral:
\begin{equation}
    \label{eq:fresnel}
    K(x) = \frac{1}{\sqrt \pi}e^{-ix^2 -i\pi/4}
    \int_x^\infty e^{i t^2} \ dt.
\end{equation}
The angles $\theta, \theta'$ are shown in Fig.~\ref{fig:angles}. 
In the half-line problem, $\theta = \pi/2$, because the incident wave
is normal to the wall.  In addition, in Eq.~(\ref{eq:uniform}) we
understand that the origin of the coordinate system is at the tip of
the half-line.

\begin{figure}
    \centerline{\epsfig{figure=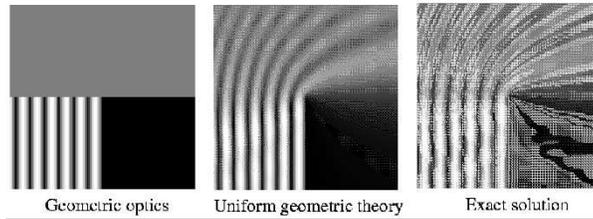,width=8cm}}
    \caption[Diffraction by a half-line: four solutions]
    { Comparison of the geometric optics approximation, the uniform
      geometrical theory, and the exact solution for the diffraction
      of quantum particle off an infinite half-line screen. The screen
      extends from the lower center to the center of the picture.
      Plotted is the norm of the wavefunction, $|\psi|$. Note the
      stark discontinuities on the shadow and reflection boundaries in
      the geometric optics approximation.  Regions of high probability
      amplitude are light.}
    \label{fig:diffraction}
\end{figure}

In Fig.~\ref{fig:diffraction} we compare the result of
Eq.~(\ref{eq:uniform}) to the exact solution for the half-line.  The
prediction of geometric optics is also shown.
There is very good agreement between the exact solution and the
uniform geometric theory.  Note especially that the discontinuities on
the shadow and reflection boundaries are completely removed by the
uniform theory.

\section{Semiclassics in the energy domain}
\label{sec:energy}
Now we turn back to the problem of calculating the transmission
properties of our resonator.  We need to find an expression for the
Green function for the resonator, because the transmission can be
easily written in terms of the diagonal part of the energy Green
function~\cite{chan97}:
\begin{equation}
    \label{eq:conductance}
    T(E) \propto {\rm Re}\left[ 
        i G({\bf r}_{\rm QPC},{\bf r}_{\rm QPC},E)\right],
\end{equation}
where ${\bf r}_{\rm QPC}$ is the center of the QPC.  The physical
reason that $G({\bf r}_{\rm QPC},{\bf r}_{\rm QPC},E)$ appears in
Eq.~(\ref{eq:conductance}) is because all waves enter our resonator
within a fraction of a wavelength of that point.  If a point source of
waves is launched at ${\bf r}_{\rm QPC}$ with energy $E$, the complex
number $G({\bf r}_{\rm QPC},{\bf r}_{\rm QPC},E)$ is just the
amplitude for returning to ${\bf r}_{\rm QPC}$.  If a significant
fraction of the rays emanating from the point source return to ${\bf
  r}_{\rm QPC}$ in phase, then $G({\bf r}_{\rm QPC},{\bf r}_{\rm
  QPC},E)$ will be appreciable, and the returning waves will beat
against the original wave and have a large effect on the transmission.

\subsection{Geometric orbits in the semiclassical propagator}
\begin{figure}
    \begin{center}
        \centerline{\epsfig{figure=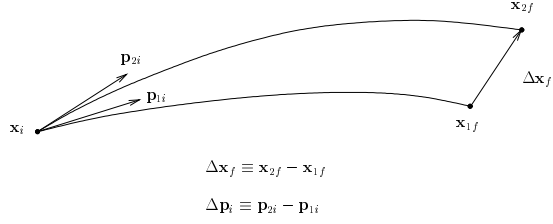,width=8cm}}
        \caption[Two trajectories]{Two trajectories launched 
          from the same point with slightly different momenta ${\bf
            p}_{1i}$ and ${\bf p}_{2i}$.  Each momentum has the same
          magnitude; only the directions are different.  The labels
          $i, f$ stand for initial and final.
          }
        \label{fig:stability}
    \end{center}
\end{figure}

The 2D semiclassical energy Green function $G_{\rm sc}({\bf r},{\bf
  r}',E)$ can be written as a sum over paths from ${\bf r}$ to ${\bf
  r}'$ thus~\cite{gutzwiller90}:
\begin{equation}
    \label{eq:sc_green}
    G_{\rm sc}({\bf r},{\bf r}',E) = \frac{2\pi}{(2\pi i)^{3/2}}
    \sum_{\rm paths} \frac{1}{\sqrt{A}}
    \exp\left[i S({\bf r},{\bf r}') - i\pi\mu/2\right],
\end{equation}
where $S({\bf r},{\bf r}')$ is the action and $\mu$ is the Maslov
index for the path. The stability coefficient $A$ is given by
\begin{equation}
    \label{eq:stability}
    A = \frac{\partial {\bf x}_{\perp f}}{\partial {\bf p}_i}
    = \lim_{\Delta{\bf p}_i\to 0}
    \frac{\Delta{\bf x}_{\perp f}}{\Delta{\bf p}_i},
\end{equation}
using the definitions from Fig.~\ref{fig:stability}, and where
$\Delta{\bf x}_{\perp f}$ is the component of $\Delta {\bf x}_f$ that
is perpendicular to ${\bf p}_{1f}$.
The coefficient $A$ describes the stability of trajectories beginning
from a particular point in phase space.  If $A$ is large, then small
changes in the initial direction of the trajectory lead to large
displacements in the final positions.  More precisely, if the distance
$\Delta{\bf x}_{\perp f}$ grows exponentially with the length of the
trajectories, we say that the trajectories are chaotic.  If
$\Delta{\bf x}_{\perp f}$ grows only linearly, then they are stable.

\begin{figure}
    \begin{center}
        \centerline{\epsfig{figure=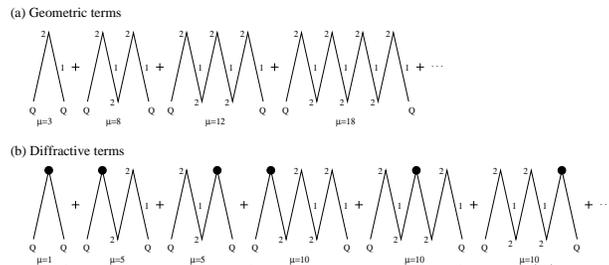,width=8cm}}
        \caption[Orbits entering the semiclassical sum]
        {Diagrams of orbits included into the semiclassical Green
          function in Eq.~(\ref{eq:sc_green}).  In (a), the horizontal
          orbits are shown.  The upper vertices represent bounces off
          the reflector, while the lower vertices signify reflections
          off the QPC.  The Maslov indices are shown for each part of
          each orbit, and the total index is given below each diagram.
          In (b), the diffractive orbits are shown.  The filled circle
          represents a diffraction event, where the term is multiplied
          by the diffraction coefficient, $D$.  After a diffractive
          event, the return path does not acquire a Maslov index of 1,
          because it is a diffracted ray; only the specularly
          reflected rays participate in the caustic.}
        \label{fig:feynman}
    \end{center}
\end{figure}

For the case of our resonator in the unstable regime, only one type of
orbit enters into the sum in Eq.~(\ref{eq:sc_green}): the horizontal
orbit.  Therefore, in order to find the contributions of the geometric
orbits to the transmission spectrum, we need only the actions ($S =
kl$, where $l$ is the length of the orbit) and stability coefficients
$A$ for the primary horizontal orbit and its repetitions.  In
addition, we need to keep track of the Maslov index for each orbit.  A
series of such orbits, together with the associated Maslov indices, is
shown schematically in Fig.~\ref{fig:feynman}(a).
In this figure, the QPC/wall is located at the lower part of each
diagram, while the reflector is located at the upper part.  Each
upward (downward) sloping line segment represents part of a trajectory
from the QPC (reflector) to the reflector (QPC).  Maslov indices of
$\mu = 2$ are indicated for points where the wave is reflected at the
wall or arc-reflector, and an index of $\mu = 1$ is acquired each time
the ray passes though the focus on its return from the reflector
toward the wall.

\begin{figure}
    \begin{center}
        \epsfig{figure=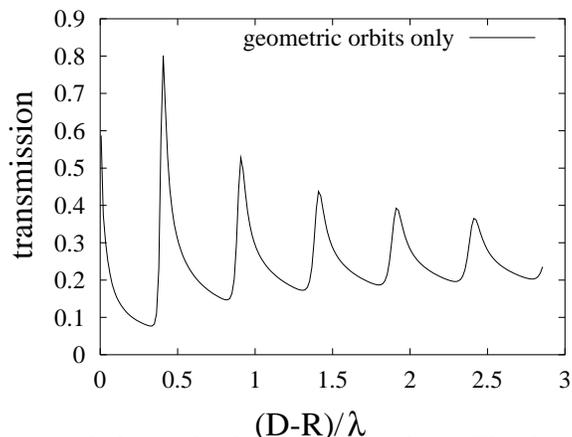,width=8cm}
        \caption[Transmission: geometric orbits]
        {Semiclassical transmission including only the geometric
          orbits.  The data is for fixed energy, corresponding to $kR
          = 40$, with the distance between the wall and reflector
          varied along the $x$-axis.}
        \label{fig:geometric}
    \end{center}
\end{figure}

Figure~\ref{fig:geometric} shows a transmission spectrum for the
resonator in the unstable regime, computed using only the horizontal
orbit.
For this calculation, the sum in Eq.~(\ref{eq:sc_green}) was cut off
after the 20th term, that is, orbits of up to 20 round trips were
included in the sum.  The half-wavelength periodicity of the spectrum
is clearly seen in the figure.  Upon comparison with
Fig.~\ref{transmission}, we see that the peak positions match very
well with the experimentally measured spectrum.  Note, however, the
absence of the peaks corresponding to diffractive orbits which are
present in Fig.~\ref{transmission}.  It is to the calculation of these
diffractive resonances that we now turn.

\subsection{Diffractive orbits in the semiclassical propagator }
So far, the Green function in Eq.~(\ref{eq:sc_green}) includes paths
that undergo evolution under the free-particle Hamiltonian, including
bounces off the wall and mirror.  In order to calculate the spectral
properties of the resonator semiclassically, diffractive orbits must
be included into the sum over paths that forms the semiclassical
propagator.  This problem has been studied by a number of
authors~\cite{vattay94,pavloff95,sieber97}.  In the literature, much
attention has been focused on finding the effects of diffraction on
the spectra of closed systems.  However, in closed systems,
diffractive orbits generally play a minor role because they are
overwhelmed by the huge number of unstable periodic orbits that do not
involve diffraction.  In this section, these methods will be extended
to include open systems.

\begin{figure}
    \centerline{\epsfig{figure=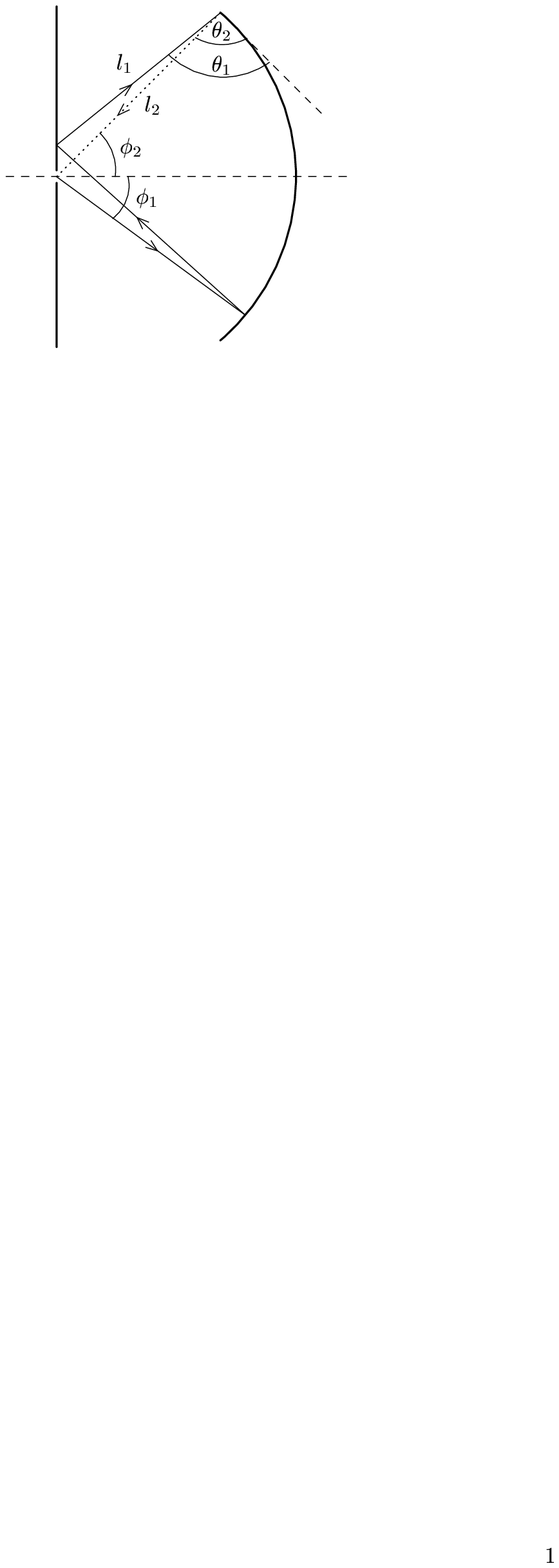,width=8cm}}
    \caption[A diffractive orbit in the semiclassical sum]
    {Here is shown the various angles relevant to a particular
      diffractive orbit entering into the semiclassical Green function
      in Eq.~(\ref{eq:sc_green_diff}).  The first leg of the orbit,
      $l_1$, is drawn with a solid line, and the second leg, $l_2$, is
      drawn with a dotted line.  This orbit corresponds to the third
      term in the sum shown in Fig.~\ref{fig:feynman}(b).}
    \label{fig:sample_orbit}
\end{figure}

For our purposes, it is sufficient to include diffraction at the level
of a single diffraction event per orbit.  Although multiply diffracted
orbits strictly belong in the semiclassical sum, in practice they can
be safely neglected.  This is because of the amplitude of an incident
ray on the reflector tip is subsequently sprayed in all directions, so
that only a small part returns in a direction that eventually leads it
back to the QPC.  Therefore, we will consider only singly diffracted
orbits.  The Green function is the product of three amplitudes: the
first for going from the starting point to the point of diffraction,
the second for the diffraction event itself, and the third for going
from the diffraction point to the final point, as follows:
\begin{eqnarray}
    \label{eq:sc_green_diff}
    G_{\rm diff}({\bf r},{\bf r}',E)&=&
    G_{\rm sc}({\bf r},{\bf r}_d,E) 
    D(l_1,l_2,\theta_1,\theta_2,E) 
    G_{\rm sc}({\bf r}_d,{\bf r}',E) \\
    \nonumber
    &=& 
    -\frac{1}{2\pi i}
    \sum_{\rm paths} 
    \frac{D(l_1,l_2,\theta_1,\theta_2,E)}{\sqrt{A_1 A_2}}
    \cos\phi_1\cos\phi_2 
    \exp\left[i (S_1+ S_1)
    - i\pi(\mu_1+\mu_2)/2
\right],
\end{eqnarray}
where the diffraction event occurs at position ${\bf r}_d$, and the
total path from ${\bf r}$ to ${\bf r}'$ is made up of two legs, one of
length $l_1$, with stability coefficient $A_1$, and the other of
length $l_2$, with stability coefficient $A_2$.  Each of these legs
has an action $S_1 = k l_1$, $S_2 = k l_2$ and Maslov index $\mu_1$,
$\mu_2$ associated with it. The factors $\cos\phi_1$, $\cos\phi_2$
represent the coupling of each leg to the QPC, and will be discussed
in the next section.  The various parameters are illustrated in
Fig.~\ref{fig:sample_orbit} for one of the shorter orbits,
corresponding to the third term in Fig.~\ref{fig:feynman}(b).  In that
figure, the first few terms entering into Eq.~(\ref{eq:sc_green_diff})
are shown, where diffraction events are represented by filled circles.

The diffraction coefficient $D(l_1,l_2,\theta_1,\theta_2,E)$ depends
on the lengths of each leg as well as the angles that the incident and
diffracted ray make with the surface at the tip of the obstacle.
These various lengths and angles are shown in
Fig.~\ref{fig:sample_orbit}.  The diffraction coefficient in the sum
above is similar to that appearing in our study of the half-line,
differing only in the argument of the Fresnel integral.  It is given
by,
\begin{eqnarray}
    \label{eq:diff_coeff2}
    D(\theta_1,\theta_2,l_1,l_2,k) &=& 
    -\mbox{sgn}(a_i)K\left(|a_i| \sqrt{\frac{k l_1 l_2}{l_1+l_2}}\right)+
    \mbox{sgn}(a_r)K\left(|a_r| \sqrt{\frac{k l_1 l_2}{l_1+l_2}}\right),\\
    \nonumber
    a_{i,r} &=& \sqrt{2}\cos\left(\frac{\theta_1\mp\theta_2}{2}\right),
\end{eqnarray}
and $K(x)$ is the modified Fresnel integral defined in Eq.~(\ref{eq:fresnel}).

\begin{figure}
    \centerline{\epsfig{figure=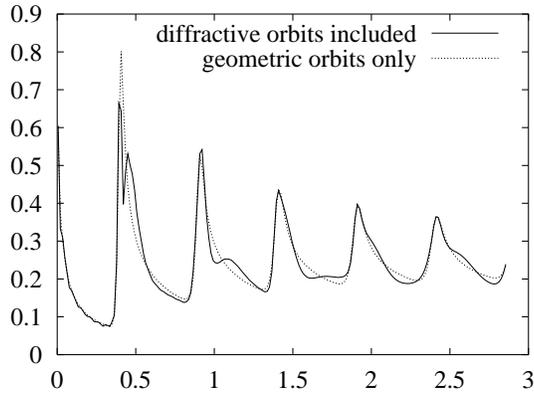,width=8cm}}
    \caption[Transmission: geometric and diffractive orbits]
    {Semiclassical transmission including both the geometric and
      diffractive orbits.  The data is for fixed energy, corresponding
      to $k R = 40$, with the distance between the wall and reflector
      varied along the $x$-axis.  The result without diffraction is
      shown with a dashed line for comparison.}
    \label{fig:geo+diff}
\end{figure}

The effect of the diffractive terms in the sum is shown in
Fig.~\ref{fig:geo+diff}.  For this calculation, all orbits with up to
20 round trips between the wall and mirror and zero or one diffractive
events were included in the sum.
We see that the effect of the diffractive terms is to modulate the
geometric result, with new peaks appearing to the right of the
geometric peaks.  The theoretical curve appearing in
Fig.~\ref{fig:geo+diff} is overlaid with the experimental data in
Fig.~\ref{transmission}; the agreement between theory and experiment
is quite good, both in the peak positions and widths of the geometric
and diffractive peaks.  We emphasize here that the semiclassical
prediction breaks down for reflector/wall separations near $D = R$,
because there the focus approaches the point where the Green function
is evaluated, and the semiclassical prediction diverges.  This is the
reason for the incorrect, large transmission calculated at $D = R$.

\begin{figure}
    \centerline{\epsfig{figure=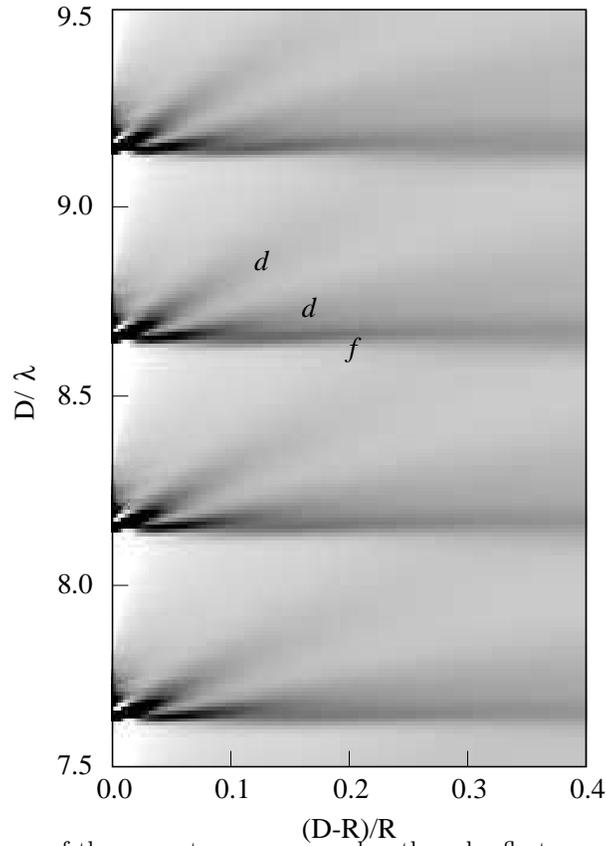,width=8cm}}
    \caption[Semiclassical transmission vs. wavelength and reflector
    separation]{Semiclassical conductance of the resonator versus
      wavelength and reflector position in the unstable regime.  The
      plot is exactly analogous to the experimental data shown on the
      right side of Fig.~\ref{pretty}.  Note that the semiclassical
      prediction breaks down near $D=R$ (see text).}
    \label{fig:sc_avoidedcrossing}
\end{figure}

In Fig.~\ref{fig:sc_avoidedcrossing}, we plot the semiclassically
calculated transmission of the resonator in the unstable regime versus
both reflector/wall separation and wavelength.  The parameters are
identical to those of the experimental data shown in
Fig.~\ref{pretty}.  The separation of the diffractive peaks from the
Fabry-Perot peaks with increasing reflector/wall separation is quite
clear.  Two diffractive peaks per geometric peak are visible.  The
half-wavelength periodicity is also apparent.  The similarity between
theory and experiment in these two figures is striking.

\subsection{Semiclassical wavefunctions}
\begin{figure}
    \begin{center}
        \epsfig{figure=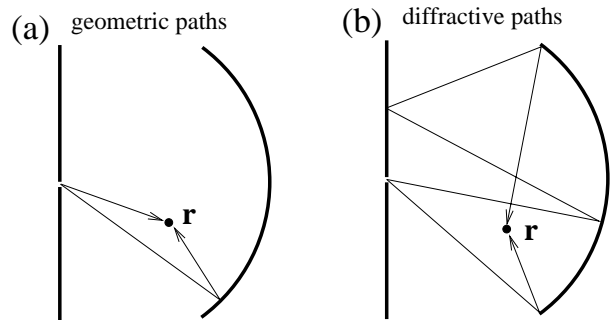,width=8cm}
        \caption{In (a), the two shortest geometric trajectories from
          the QPC to the point ${\bf r}$ are shown.  Such orbit are
          specularly reflected off the wall and mirror.  In (b), the
          two shortest diffractive trajectories from the QPC to the
          point ${\bf r}$ are drawn.  }
        \label{fig:short_traj}
    \end{center}
\end{figure}

In this section, we describe the procedure for including diffraction
into a semiclassical calculation of the resonator wavefunctions.  As
in the previous sections, we can split up the sum over paths into a
geometric part and a diffractive part.  The contribution to $\psi$
from the geometric orbits is proportional to the semiclassical
amplitude for getting from the QPC to the point of interest:
\begin{equation}
    \psi_{\rm geo}({\bf r}) \propto G_{\rm sc}({\bf r_{\rm QPC}},{\bf r},E),
\end{equation}
where ${\bf r}_{\rm QPC}$ is the location of the QPC, and ${\bf r}$ is
the location of interest.  This Green function is the same as we have
already encountered in the semiclassical expression for the
transmission neglecting diffraction, appearing in
Eqs.~\ref{eq:conductance} and~\ref{eq:sc_green}.  The only difference
is that now we are looking at an off-diagonal element of the Green
function, rather than a diagonal element.  Some of the shorter
trajectories that are included in the sum are shown in
Fig.~\ref{fig:short_traj}(a).  All such trajectories begin at the QPC
and end at the point ${\bf r}$, undergoing specular reflection at the
wall and reflector.

\begin{figure}
    \centerline{\epsfig{figure=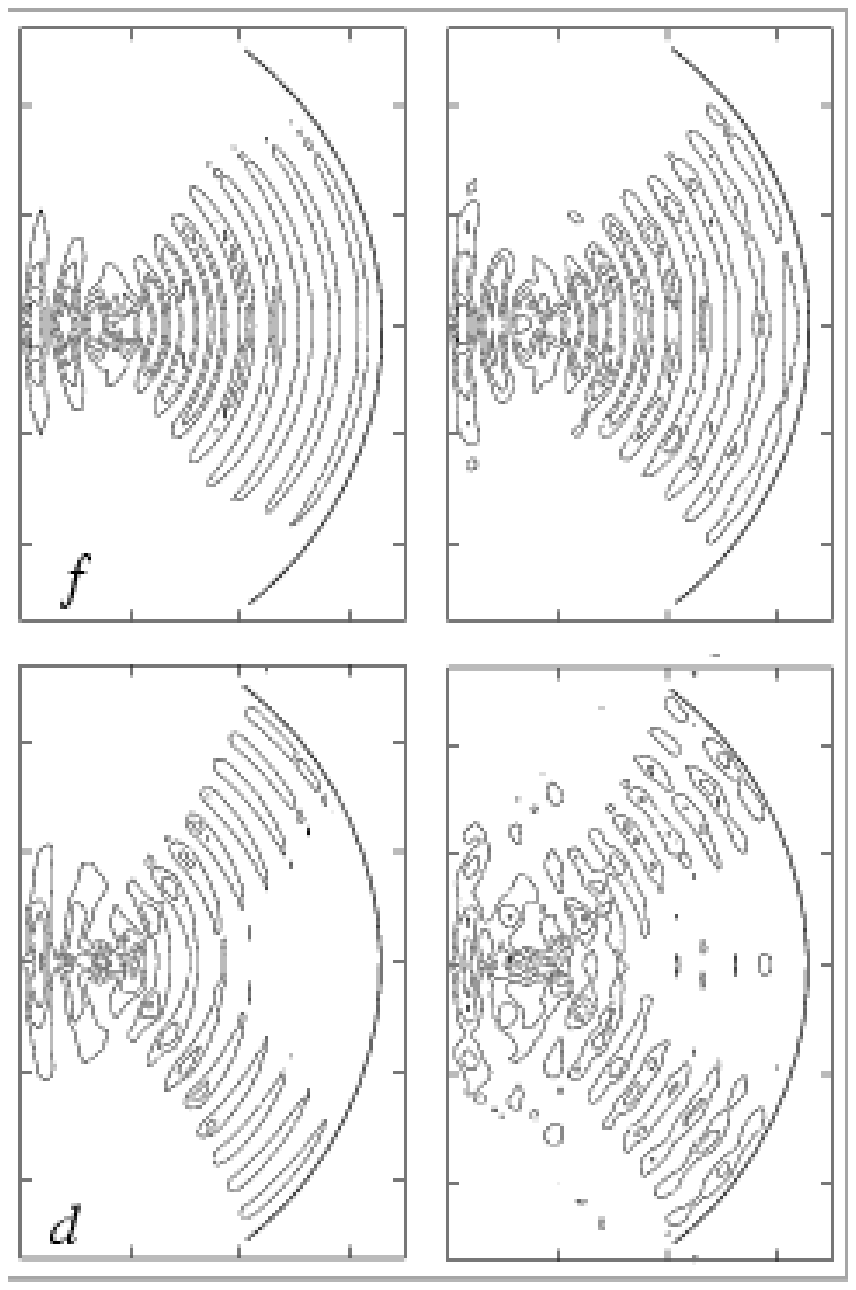,width=8cm}}
    \caption{Two wavefunctions calculated semiclassically.   On the
      left, only the geometric paths were included in the sum; on the
      right trajectories that diffract off the edges of the reflector
      were taken into account.  These plots are for identical
      parameter sets as those shown in Fig.~\ref{wavefunctions}.}
    \label{fig:wavefunctions_sc}
\end{figure}

To include diffraction, we simply add diffractive terms to the sum,
much as we did to incorporate the effects of diffraction in the
formula for the Green function in Eq.~\ref{eq:sc_green_diff}.  We have
\begin{equation}
    \label{eq:wf_sc}
    \psi({\bf r})\propto
    G_{\rm sc}({\bf r}_{\rm QPC},{\bf r},E)+
    G_{\rm sc}({\bf r}_{\rm QPC},{\bf r}_{\rm tip},E) 
    D(l_1,l_2,\theta_1,\theta_2,E) 
    G_{\rm sc}({\bf r}_{\rm tip},{\bf r},E),
\end{equation}
where the diffraction coefficient $D(l_1,l_2,\theta_1,\theta_2,E)$ is
identical to that appearing in Eq.~\ref{eq:diff_coeff2}.  The lengths
$l_1,l_2$ and angles $\theta_1,\theta_2$ are defined the same way as
in Fig.~\ref{fig:sample_orbit}, with the obvious difference that the
final point of the trajectory is no longer the location of the QPC,
but the location of interest, ${\bf r}$.  In
Fig.~\ref{fig:wavefunctions_sc}
we show the result of a semiclassical calculation for the
wavefunctions pictured in Fig.~\ref{wavefunctions}.  For comparison,
we also show the result when diffractive paths are left out of the
sum.  In the calculation, all singly diffracted paths involving up to
20 bounces were included.

We saw earlier that diffraction has a large effect on the transmission
spectrum, which in turn could be derived from the value of the
wavefunction near the QPC.  By contrast, we see now that the inclusion
of diffraction has a relatively minor effect on the overall appearance
of the wavefunctions.  The explanation for this apparent paradox is
that the QPC is near many reflection boundaries, where the diffractive
corrections are especially large.

The transmission through a small aperture (here the QPC) is extremely
sensitive to small returning bits of amplitude, which interfere
coherently with the wave entering through the aperture to modulate the
transmission.  In our case, a major source of returning amplitude is
provided by the diffractive orbits.  This is analogous to scanning
tunneling microscope ``quantum corral'' imaging~\cite{corral-refs},
where the tunneling from the tip plays the role of the quantum point
contact, and reflections from atoms and impurities represent the
diffraction and geometric scattering off the ends of the reflector.
The modulation of the transmission through the QPC by the presence of
the reflector can be understood in terms of a small returning wave
amplitude beating against a much stronger ``nascent'' amplitude coming
out of the QPC: if the amplitude from the QPC in absence of the
reflector is $A$, and the returning amplitude with the reflector
present is $a$, then the total amplitude at the QPC is simply $A+a$.
The transmission will be proportional to the square of this amplitude,
\begin{equation}
   T \propto |A+a|^2 = |A|^2 + aA^* +a^*A + |a|^2.  
\end{equation}
The two interference terms above, linear in $a$, are responsible for
all the structure in the transmission when the reflector is present
in the cavity. 

\section{Semiclassics in the time domain}
\label{sec:time}
\begin{figure}
    \centerline{\epsfig{figure=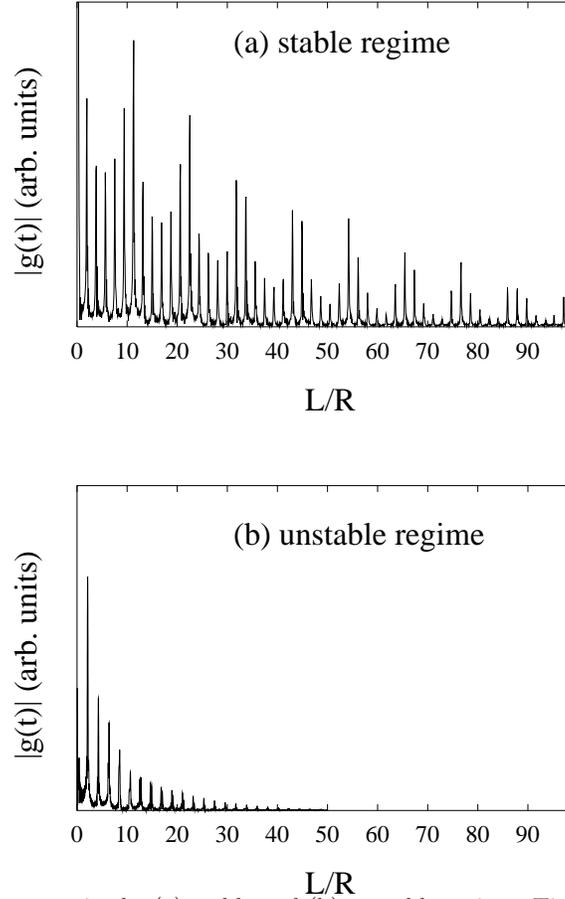,width=8cm}}
    \caption[Return spectra]{
      Experimental return spectra in the (a) stable and (b) unstable
      regime. Time has been converted to the ratio $L/R$, where $L$ is
      the length of the orbit.  For these plots the opening angle was
      $115^\circ$, and the reflector-wall separation was 28.5~cm and
      32.5~cm, respectively.  For the unstable case, the theoretically
      predicted envelope of the decay given by the Lyapunov exponent
      is shown as well.}
    \label{fig:fourier1}
\end{figure}

Further evidence of diffractive orbits in the transmission spectrum
can be obtained by analyzing the spectrum in the time domain.  The two
representations are related by the Fourier transform:
\begin{equation}
    \label{eq:fourier}
    g(t) = \int_{-\infty}^\infty S_{11}(\omega)e^{i\omega t} \ dt.
\end{equation}
Here $g(t)$ represents the amplitude for a pulse launched from the QPC
at time $t=0$ to return at time $t$.  That is, if a short pulse were
emitted from the antenna at time $t = 0$, echos would return to the
antenna at certain later times.  These echos are indicated by peaks in
the return spectrum.  In Fig.~\ref{fig:fourier1}, we plot the
amplitude $|g(t)|$ as a function of time for the resonator in the
stable and unstable regimes. In these plots, time has been normalized
so that one unit is the time it takes for light to travel a
distance equal to the radius of curvature of the reflector.
In the stable regime, the echos persist for hundreds of bounces,
indicating that indeed the dynamics is stable in this regime.  In this
regime, the lifetime of the states is limited by resistive losses in
the copper plates of the resonator, which was quite small: typical
quality factors of the resonances in this regime were $Q \sim 3000$.
However, in the unstable regime, the echos are significantly reduced
in amplitude after only a few returns.  For the first few return
peaks, $|g(t)|^2$ decays exponentially with a decay constant given
approximately by the Lyapunov exponent for the horizontal orbit.
Later peaks, where diffractive orbits are more important, also decay
exponentially but with a different decay constant.

\begin{figure}
    \centerline{\epsfig{figure=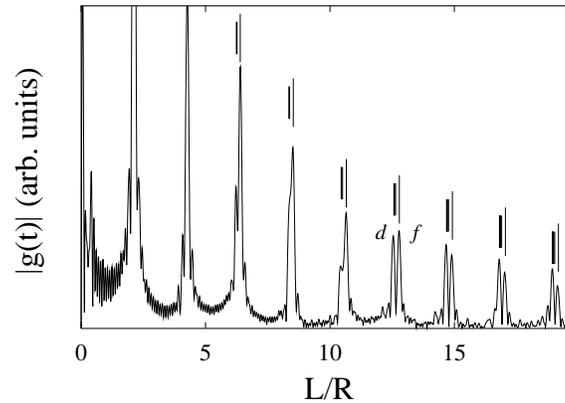,width=8cm}}
    \caption[Return spectra: unstable regime]{
      Experimental return spectra for the unstable regime. Time has
      been converted to the ratio $L/R$, where $L$ is the length of
      the orbit.  The splitting of the peaks clearly demonstrates the
      influence of both Fabry-Perot type orbits (marked $f$) and
      diffractive orbits (marked $d$), which are slightly shorter.
      The calculated lengths of the orbits are shown by vertical bars;
      short bars for the diffractive orbits, and longer bars for the
      horizontal orbit.  For these plots the opening angle was
      $115^\circ$, and the reflector/wall separation was 32.5~cm.}
    \label{fourier}
\end{figure}

In Fig.~\ref{fourier}, we show an expanded view of some of the return
peaks shown in Fig.~\ref{fig:fourier1}(b).  Of importance here is the
splitting of the return peaks which is visible on echos 5-9.  This
splitting is due to the coexistence of orbits with slightly different
periods.  The longer of these orbits is just the horizontal orbit,
which appears on the right of each group.  The left peak of each group
is made up of a family of diffractive orbits of nearly the same
length.  We have done a quantitative study of the lengths of the
closed orbits and find excellent agreement with the observed
splitting.  The calculated lengths of the orbits appear in the plot as
vertical bars above the peaks.  The horizontal orbit length is marked
with a longer bar.  The lengths of all the orbits in units of the
radius of curvature appear in Table.~\ref{length_table}.  The presence
of this splitting in the return spectrum is strong evidence in support
of the claim that diffraction off the edges of the reflector supports
other closed orbits, which lead to resonances in the transmission
spectra.  Note that for the long orbits, the diffractive peaks are
even stronger than the peaks from the geometric orbit.  This is
because the number of diffractive orbits increases linearly with the
length of the orbit, whereas there is always only one geometric orbit,
regardless of length.

\begin{table}
    \begin{center}
        \newcommand{\rb}[1]{\raisebox{1.5ex}[0pt]{#1}}
\begin{tabular}{|c|c|c||c|c||c|c||c|}
    \hline
    \multicolumn{8}{|c|}{\rule[-3mm]{0mm}{8mm}
      \bfseries Lengths of diffractive orbits}\\ 
    \hline 
    $N$ & $N_1$&$N_2$&$L_1$&$L_2$&$L_{\rm tot}$&
    $L_{\rm hor}$& $\Delta$\\
    \hline \hline
    2&1&1& 1.03658 & 1.03658 & 2.07316 & 2.13114&0.0580\\ \hline\hline
    4&3&1& 3.12029 & 1.03658 & 4.15687 & 4.26228&0.1054\\ \hline\hline
    &3&3&3.12029&3.12029& 6.24058& &0.1528\\ \cline{2-6} \cline{8-8}
    \rb{6}&5&1&5.22441&1.03658&6.26099&\rb{6.39342}&0.1324\\ \hline\hline
    &5&3&5.22441&3.12029&8.34470& &0.1799\\ \cline{2-6}\cline{8-8}
    \rb 8&7&1&7.34315&1.03658&8.37973&\rb{8.52456}&0.1448\\ \hline\hline
    &5&5&5.22441&5.22441&10.4488& &0.2069\\ \cline{2-6} \cline{8-8}
    10&7&3&7.34315&3.12029&10.4634&10.6557&0.1923 \\ \cline{2-6} \cline{8-8}
    &9&1&9.47048&1.03658&10.5071& &0.1486\\ \hline\hline
    &7&5&7.34135&5.22441&12.5677& &0.2193\\ \cline{2-6} \cline{8-8}
    12&9&3&9.47048&3.12029&12.5908&12.7868&0.1961 \\ \cline{2-6} \cline{8-8}
    &11&1&11.6009&1.03658&12.6375& &0.1494\\ \hline\hline
    &7&7&7.34315&7.34315&14.6863& &0.2317\\ \cline{2-6} \cline{8-8}
    &9&5&9.47048&5.22441&14.6949& &0.2231\\ \cline{2-6} \cline{8-8}
    \rb{14}&11&3&11.6009&3.12029&14.7212&\rb{14.9180}&0.1968\\ \cline{2-6} \cline{8-8}
    &13&1&13.7309&1.03658&14.7675& &0.1505\\ \hline\hline
    &9&7&9.47048&7.34315&16.8136& &0.2355\\ \cline{2-6} \cline{8-8}
    &11&5&11.6009&5.22441&16.8253& &0.2238\\ \cline{2-6} \cline{8-8}
    \rb{16}&13&3&13.7309&3.12029&16.8512&\rb{17.0491}&0.1979\\ \cline{2-6} \cline{8-8}
    &15&1&15.8622&1.03658&16.8988& &0.1503\\ \hline\hline
    &9&9&9.47048&9.47048&18.9410& &0.2393\\ \cline{2-6} \cline{8-8}
    &11&7&11.6009&7.34315&18.9441& &0.2362\\ \cline{2-6} \cline{8-8}
    18&13&5&13.7309&5.22441&18.9553&19.1803&0.2250 \\ \cline{2-6}\cline{8-8} 
    &15&3&15.8622&3.12029&18.9825& &0.1978\\ \cline{2-6}\cline{8-8} 
    &17&1&17.9946&1.03658&19.0312& &0.1491\\ \hline\hline
    &11&9&11.6009&9.47048&21.0714& &0.2400\\ \cline{2-6}\cline{8-8} 
    &13&7&13.7309&7.34315&21.0741& &0.2373\\ \cline{2-6}\cline{8-8} 
    20&15&5&15.8622&5.22441&21.0866&21.3114&0.2248\\ \cline{2-6}\cline{8-8} 
    &17&3&17.9946&3.12029&21.1149& &0.1965\\ \cline{2-6} \cline{8-8}
    &19&1&20.1231&1.03658&21.1597& &0.1517\\ \hline\hline
    &11&11&11.6009&11.6009&23.3018& &0.2407\\\cline{2-6}\cline{8-8} 
    &13&9&13.7309&9.47048&23.2014& &0.2411\\ \cline{2-6}\cline{8-8} 
    &15&7&15.8622&7.34315&23.2054& &0.2371\\ \cline{2-6}\cline{8-8} 
    \rb{22}&17&5&17.9946&5.22441&23.2190&\rb{23.4425}&0.2235\\ \cline{2-6}\cline{8-8} 
    &19&3&20.1231&3.12029&23.2434& &0.1991\\ \cline{2-6} \cline{8-8}
    &21&1&22.2548&1.03658&23.2914& &0.1511\\ \hline
\end{tabular}
    \end{center}
    \caption[Lengths of diffractive orbits]
    {Lengths of the diffractive orbits. The columns are: $N$ = number
    of half-bounces in orbit; $N_1$ = number of half-bounces in first
    leg; $N_2$ = number of half-bounces of second leg; $L_1$ = length
    of first leg; $L_2$ = length of second leg; $L_{\rm tot} = L_1
    +L_2$; $L_{\rm hor}$ = length of horizontal orbit; $\Delta =
    L_{\rm hor}-L_{\rm tot}$.  The configuration of the resonator for
    this data was $D = 1.0656R$, $\alpha = 115^\circ$.  All numbers
    are given in terms of the radius of curvature of the reflector,
    which is taken to be unity.}
    \label{length_table}
\end{table}

\section{The avoided crossing}
\label{sec:avoided}
We now turn our attention to the avoided crossing that appears in
Fig.~\ref{pretty} near $(D-R)/R = 0$.  We mentioned earlier that the
level repulsion at this point is in part mediated by diffraction.  To
show this, we compare our open resonator with three closely-related
closed systems.

\begin{figure}
    \centerline{\epsfig{figure=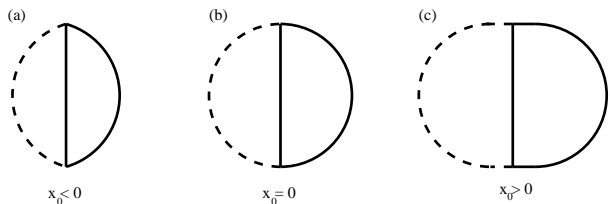,width=8cm}}
    \caption[Lemon, circle, and stadium billiards]
    {The lemon (a), circle (b), and stadium (c) billiards.  We are
      interested only in states of these billiards which are even
      about the $x$-axis, and odd about the $y$-axis.}
    \label{lemon}
\end{figure}

In the stable regime, imagine closing the QPC aperture and increasing
the open angle of the arc-shaped reflector until it touches the
straight wall.  The resulting shape is one-half of a lemon billiard
(see Fig.~\ref{lemon}(a)).  The shape of the lemon billiard is
determined by the parameter $x_0 \equiv (D - R)/R < 0$, which is the
$x$-value of the center of curvature of the arc on the right in units
of the radius of curvature.  The lemon billiard has been studied
before~\cite{katine97,postmodern}; for our purposes it is important
that the classical dynamics in the lemon billiard is dominated by a
large regular region.

As $x_0$ is increased to $0$, the area enclosed by the arc becomes
half of a perfect circle, and the closed system becomes completely
integrable (see Fig.~\ref{lemon}(b)).  The eigenstates of a circular
billiard are $J$-type Bessel functions.

Finally, as $x_0$ is made positive, the semicircular arc can be
extended with horizontal straight segments to form half of a stadium
billiard.  The classical dynamics in the stadium billiard is
completely chaotic.  See Fig.~\ref{lemon}(c).

The eigenstates of these three systems share certain properties with
the scattering states of our open system.  In these closed systems,
there are no wall ends so we expect diffraction to play a smaller role
in the energies and wavefunctions.

The similarity between these closed systems and our resonator is
greatest in the stable regime.  In this regime, the states of the
resonator are essentially the same as the states of the lemon billiard
that are even about the $x$-axis, and odd about the $y$-axis.  (The
straight wall in the resonator lies on the $y$-axis, thus enforcing a
node there, while the symmetric position of the antenna means that it
only excites states that are even about the $x$-axis.)  As the
reflector position is varied, we correspondingly change the parameter
$x_0 < 0$ of the lemon billiard, and compare the states of each
system.  In the experiment, we see an avoided crossing at $x_0 = 0$.
To investigate this, we can follow the states of the lemon/stadium
billiard as the parameter $x_0$ is swept slowly through zero.

\begin{figure}
    \centerline{\epsfig{figure=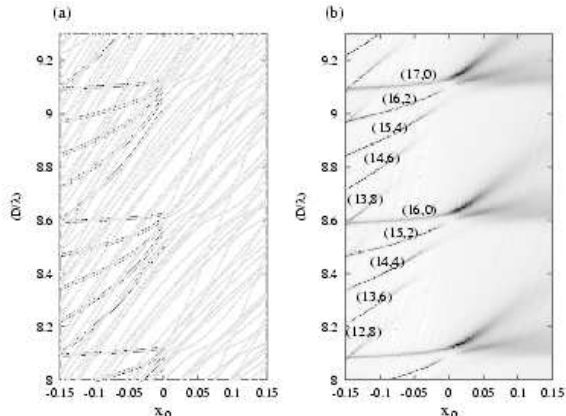,width=8cm}}
    \caption[Levels of the lemon/stadium billiard]
    {On the left, the states of a closed billiard are shown, as the
      parameter $x_0$ is varied.  For $x_0 < 0$, the billiard is a
      lemon billiard; for $x_0 = 0$, it is a circle; for $x_0 > 0$ it
      is a stadium.  For $x_0 < 0$, we have darkened the lines
      corresponding to the allowed states of the corresponding open
      system (see Fig.~\ref{fig:lemonstates} for the corresponding
      eigenfunctions).  On the right, the experimental transmission
      spectra for the open system are shown for the same parameter
      range.  Note the similarity of the figures in the range $x_0 <
      0$, where the open system exhibits stable classical dynamics.}
    \label{fig:level_diagram}
\end{figure}

In Fig.~\ref{fig:level_diagram}(a)
we plot the (even-about-$x$, odd-about-$y$) levels of the
lemon/stadium billiard as a function of the parameter $x_0 = (D-R)/R$
and wavelength.  On the right, the corresponding experimental
transmission spectra for the open system are shown.  Notice the exact
matching of the transmission peaks with the lemon billiard states in
the stable regime.  This excellent agreement is an artifact of the
fact that, while our resonator is a geometrically \emph{open} system,
it is classically \emph{closed}, in the sense that almost all
trajectories beginning at the QPC that hit the reflector are doomed to
forever remain in the region between the wall and the reflector.  We
only begin to notice the ``openness'' of our resonator for the peaks
corresponding to large numbers of angular nodes. As we shall see, such
states have a large angular spread, and it grows with decreasing
$x_0$, so that eventually the caustics of the classical orbits
supporting these states approach the tip of the reflector.  When this
happens, the states broaden and disappear.  For example, this is
apparent for the peak with quantum numbers (13,8) at around $x_0 =
-0.1$.  The wavefunction corresponding to this peak has 13 radial
nodes between the $y$-axis and the reflector, and 8 angular nodes.
The states with fewer angular nodes do not have such a wide angular
extent, and therefore they do not disappear from the spectra until the
reflector is farther from the wall.

It is apparent that there are many states in the closed lemon billiard
that do not appear at all in the analogous open system.  All such
states have considerable amplitude near the walls of the lemon
billiard that are `missing' from the open resonator---therefore in the
open system the analogous states escape the resonator immediately.
Thus they are absent in the experimental transmission spectra.

At $x_0 = 0$, there is an avoided crossing in both the open and closed
systems.  However, we see that the level repulsion is stronger in the
open system.  For example, observe the level repulsion of the states
(17,0) and (16,2).  The distance of closest approach of these two
levels is four times greater in the open system than in the closed
system.  (For the open system, we judge the distance between the
``levels'' as the distance between the peak maxima.)  Now, the major
difference between the two systems as far as the eigenstates are
concerned is the presence of diffractive orbits in the open case.
Semiclassically speaking, there are diffractive paths that go from one
state to the other and introduce a coupling that increases the level
splitting.  Therefore, we believe that the avoided crossing is in
large part mediated by \emph{diffraction} in the open system.

For $x_0 > 0$, the correspondence between the closed and open systems
comes to a somewhat abrupt halt.  This is apparent in
Fig.~\ref{fig:level_diagram}.  The reason is as follows.  In the
closed system, orbits with times up to about the Heisenberg time, $t_H
\equiv 1/\!\Delta E$, affect the properties of the quantum states.  In
the open system, on the other hand, only a few orbits from a very
specific part of phase space stay in the system for more than a few
bounces.  Therefore, we can expect states of the closed and open
systems to correspond only if a significant subset of trajectories
remains in the non-escaping region of phase space for roughly the
Heisenberg time.  Trajectories that leave that region of phase space
explore parts of phase space where the systems differ and the
correspondence breaks down.  For the circle billiard at our
experimental parameters, the particle makes approximately $7$
horizontal bounces within the Heisenberg time of the quarter-stadium.

Using the 2D density of states, we can estimate the range of $x_0$ for
which we may expect rough correspondence between the eigenstates of
the closed system with the resonances of the open system in the
unstable regime.  There is another characteristic time, $t_\lambda$,
which we will call the Lyapunov time, which is the characteristic time
that it takes for a trajectory to ``fall off'' the horizontal periodic
orbit.  We expect correspondence when $t_H \lesssim t_\lambda$.  The
Lyapunov time is
\begin{equation}
    \label{eq:lyapunov}
    t_\lambda = \frac{L}{k\lambda_{\rm Lyap}},
\end{equation}
where $L$ is the length of the periodic orbit, $\lambda_{\rm Lyap}$ is
the Lyapunov exponent of the orbit, and $k$ is the wavenumber ($L/k$
is the period of the orbit).  We will focus on the horizontal orbit
for the case where the system is just barely unstable, so that
$L\approx 2R$.  The Lyapunov exponent for the horizontal orbit is
given by the logarithm of the largest eigenvalue of the monodromy
matrix, linearized about the horizontal orbit,
\begin{equation}
    \label{eq:monodromy}
    M = \left(\begin{array}{cc}
            1+2x_0 & 2x_0(1+x_0)\\
            2 & 1+2x_0\\
        \end{array}\right)
\end{equation}
\cite{edwardsthesis}.  The largest eigenvalue of this matrix is given
by
\begin{eqnarray}
    \label{eq:eigenvalue}
    \nonumber
    m_+ &=& 1+2x_0 +2\sqrt{x_0(1+x_0)}\\
    &\approx& 1+2\sqrt{x_0},    
\end{eqnarray} 
for small $x_0$.  Thus the Lyapunov exponent is
\begin{eqnarray}
    \nonumber
    \lambda_{\rm Lyap} &\approx& \ln(1+2\sqrt{x_0})\\
    \nonumber
    &\approx&2\sqrt{x_0}.
\end{eqnarray}
Therefore, the Lyapunov time for the horizontal orbit is approximately
\begin{equation}
    t_\lambda \approx \frac{R}{k\sqrt{x_0}}.
\end{equation}
This must be compared with the Heisenberg time, which is $t_H = A/2\pi
= R^2/8$ (the effective area of the billiard is $A=\pi R^2/4$ because
of symmetry---we only take one quarter of the area of a full circle).
We expect the correspondence between the open and closed systems to
break down when these times are of the same order.  That is, we look
for the value of $x_0$ for which $t_H \approx t_\lambda$, which is
\begin{equation}
    x_0 \approx \left(\frac{8}{kR}\right)^2.
\end{equation}
Now, in our energy range $kR \approx 18\pi$.  Thus we have for the
value of $x_0$ at which correspondence ceases to hold,
\begin{equation}
    x_0 \approx \left(\frac{8}{18\pi}\right)^2
    \approx 0.02 \ll {\lambda\over R}.
\end{equation}
As indicated, this value of $x_0$ is much less than a wavelength,
scaled to the radius of the billiard.  This means that the states of
the closed billiard already have mixed through a number of avoided
crossings by the time the circle has been ``stretched'' by one
wavelength.  Indeed in Fig.~\ref{fig:level_diagram} we see that the
first avoided crossings are happening already near $x_0 = 0.02$,
consistent with our rough estimate.

\begin{figure}
    \centerline{\epsfig{figure=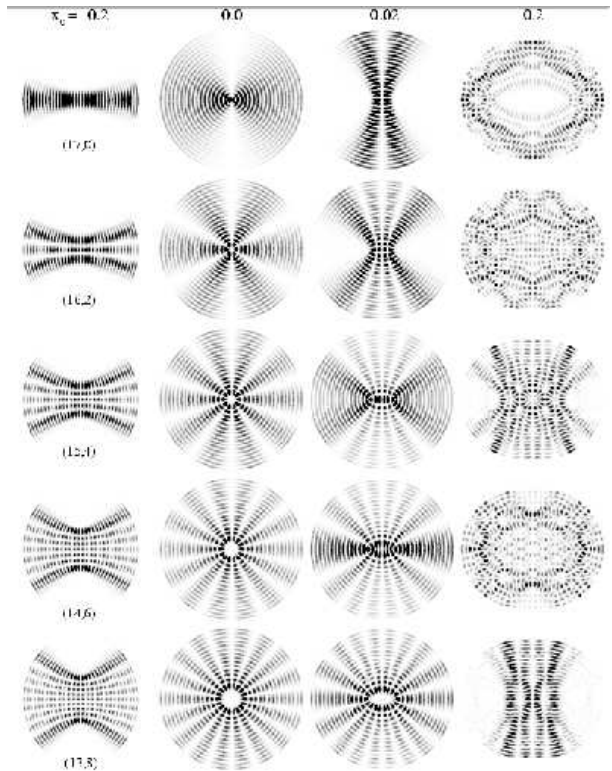,width=8cm}}
    \caption[States of the lemon/stadium billiard]
    {Eigenfunctions of the lemon/stadium billiard.  On the top line is
      shown the parameter $x_0$, which describes how far the billiard
      is from a circle.  On the left, we show some lemon billiard
      eigenfunctions, together with their quantum numbers.  To the
      right, we show the eigenfunctions associated with each lemon
      state, as the parameter $x_0$ is varied adiabatically to
      $x_0=0,0.02,0.2$.}
    \label{fig:lemonstates}
\end{figure}

A sampling of states for the closed system of the lemon/stadium
billiard is shown in Fig.~\ref{fig:lemonstates}.  In this figure, the
states corresponding to the series of quantum numbers $(17,0)$ to
$(13,8)$ are shown, as the transition from a lemon to a stadium is
made.  The corresponding values of $k$ for each state shown are given
in Table~\ref{table:lemonstates}.  The value of $x_0$ is given at the
top of each column.  We see that the angular extent of the lemon
billiard states increases with the number of angular nodes, as
mentioned previously.  As $x_0$ approaches zero, the angular extent of
each state is fully developed and covers the entire billiard, as is
required by the rotational symmetry of the circle billiard states.
The transition from the first to the second column is {\em diabatic},
that is, from left to right we track the states with the same
character.

In the third column, we have plotted the states after $x_0$ has been
{\em adiabatically} increased to the small value $x_0 = 0.02$, such
that they still bear some resemblance to the states of the circle.
This parameter value was chosen to be just before the first avoided
crossing in the closed system, in accordance with the discussion
above.  This effectively means that the periodic orbits are not too
different from those of a circle billiard; i.e., the periodic orbits
are only weakly unstable.  This means that one may still draw
analogies between the closed and open systems in this regime, although
the character of a particular resonance in the open system may be
shared by several states in the corresponding closed system.  For
example, note that the first two states in the third column have
angular lobes which are qualitatively similar to the diffractive state
shown in Fig.~\ref{wavefunctions}.  On the other hand, the third and
fourth states in that column have amplitude running from the center
out along the $x$-axis, qualitatively similar to the Fabry-Perot type
state in Fig.~\ref{wavefunctions}.  There is even the appearance of a
focus just to the right and left of the center of the billiard in
these two states, analogous to the focus seen in the Fabry-Perot state
for the open system.

In the last column, we plot the five wave functions after the
parameter $x_0$ has been adiabatically increased to $x_0 = 0.2$, or
20\% of the radius of curvature.  Now we see that the states no longer
have anything to do with the states of the open system, exactly for
the reasons given above.

\begin{table}
    \begin{center}
        \begin{tabular}{|c|c|c|c|c|}
\hline
\multicolumn{5}{|c|}{\rule[-3mm]{0mm}{8mm}
  \bfseries Eigenstates of the closed system}\\ 
\hline 
$(n,m)$ & $x_0 = -0.2$ & $x_0 = 0.0$ & $x_0 = 0.02$ & $x_0 = 0.2$ \\
\hline
(17,0) & 71.3806 & 57.3275 & 56.8273 & 51.8407 \\
(16,2) & 70.2283 & 57.2577 & 56.4768 & 51.6105 \\
(15,4) & 69.0855 & 57.1173 & 56.2240 & 51.3391 \\
(14,6) & 67.9527 & 56.9052 & 56.1205 & 50.2602 \\
(13,8) & 66.8308 & 56.6196 & 55.8705 & 50.7444 \\ \hline
\end{tabular}

    \end{center}
    \caption[Eigenenergies of the lemon/stadium billiard]
    {Eigenvalues $k$ for the  states shown in
      Fig.~\ref{fig:lemonstates}.  The quantum numbers in the first
      column are only good when $x_0$ is negative or a small positive
      number.  In the stadium regime, the classical dynamics is
      chaotic and there are no longer any good quantum numbers; hence
      the quantum number assignment for $x_0 = 0.2$ is only an
      indication of its adiabatic ancestor for smaller values of
      $x_0$.  Note also that in the quantum numbering system, we do
      not count the angular node along the $y$-axis.  This is to keep
      the analogy with the open system, where that node is trivial as
      it is forced by the position of the wall.}
    \label{table:lemonstates}
\end{table}

\section{Imaging wavefunctions with a coarse probe}
\label{sec:imaging}
We end this paper with an interesting discovery which was made in the
course of carrying out the measurements described above: namely, the
possibility of measuring an electronic wavefunction in a 2DEG.  The
viability of measuring pure state wavefunctions in the context of
microwave billiard systems has been demonstrated by many
authors~\cite{gokirmak98,sridhar92,stein92,stockman90}.  However, the
imaging of a wavefunction in a {\em real} quantum system has not yet
been achieved.

\begin{figure}
  \centerline{\epsfig{figure=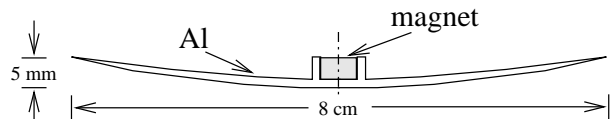,width=8cm}}
  \caption[A coarse probe]
  {\label{probe} A cross section down a diameter of the coarse probe
    used for measuring the microwave field.  It has cylindrical
    symmetry about the center axis.}
\end{figure}

We believe a technique similar to the method used here could be
applied in clean mesoscopic systems to obtain images of two
dimensional wavefunctions in a 2DEG.  In such systems, the fermi
wavelength $\lambda$ of the electrons is on the order of 50 \AA.  In
analogy to the steel ball used in the microwave experiments, an AFM
tip held close to the surface of the heterostructure could serve as a
perturbation to measure the frequency shifts.  However, the
perturbation due to a nearby AFM tip on the electron wavefunction is a
smooth potential disturbance 50-100 \AA\ in size, comparable to or
greater than $\lambda$.  One might expect that it would be difficult
to measure the nodal lines of a wavefunction with such a coarse probe.
To investigate this problem, we tried measuring a microwave mode with
a coarse probe.  Instead of the small steel ball, we used a probe 8~cm
in size, corresponding to 1.5 wavelengths, for these measurements.
The probe is shown in Fig.~\ref{probe}.  It approximately a ``chord''
of a sphere.  The probe could be moved around the upper surface of the
top plate of the cavity by means of an external magnet, in a manner
similar to that used for the steel ball perturbation.  For a
discussion of the form of the perturbation caused by this probe, see
Appendix~\ref{app:coarse-probe}.

\begin{figure}
  \centerline{\epsfig{figure=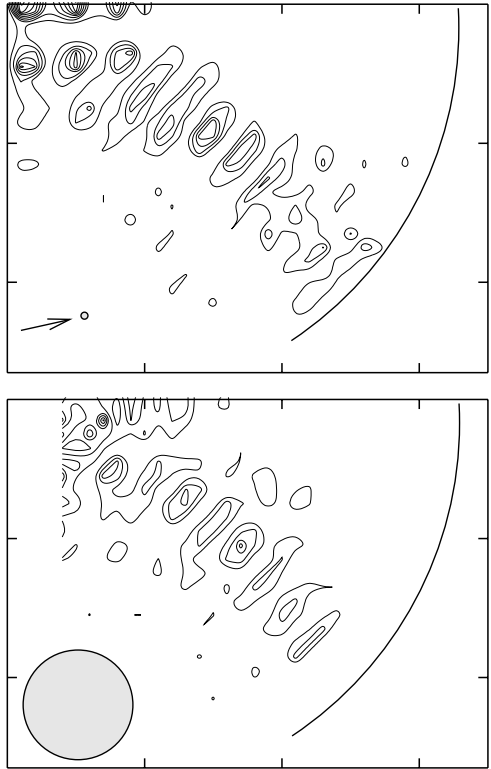,width=8cm}}
  \caption[A wavefunction measured by the coarse probe]
  {\label{fig:compare} A comparison between fine and coarse
    measurement probes.  The sizes of the probes are indicated by
    circles in the lower left corner of each graph.  In the upper
    plot, an arrow marks the probe.  The tics on all axes are 10 cm
    apart.}
\end{figure}

The results of a wavefunction measurement of the peak {\em d1} with
the coarse probe are displayed in Fig.~\ref{fig:compare}.
Only half of the wavefunction was measured.  For comparison, the same
state as measured by the 4 mm ball is shown as well.  Note that far
from the QPC, the nodal lines in the two measurements match very well.
For the coarse measurement, data near the wall and mirror were
unavailable because of the large size of the probe.  The level of
detail obtained with the coarse probe is surprising, considering that
it was over {\em three} half-wavelengths in diameter.  This result
suggests it might be possible to directly image an electron
wavefunction in a 2DEG.  A perturbative analysis of the effect of the
coarse probe is presented in Appendix~\ref{app:coarse-probe}.

\section{Conclusion}
\label{sec:conclusion}
In summary, we have demonstrated the existence of diffractive orbits
in an open microwave billiard, which give rise to resonances and
wavefunctions that would not be predicted by a simple semiclassical
theory.  Such orbits are of importance in open, unstable systems where
the number of unstable classical periodic orbits is small.  In such
systems, diffraction can play a major role in determining the spectrum
of the system.  Furthermore, we have shown that it may be possible to
measure the pure state wavefunction of an electron in a 2DEG, by using
a a coarse AFM tip as a probe.

This work was supported by NSF Grant CHE9610501.  We are grateful to the
Hewlett Packard Corporation for the loan of a network analyzer that
was used in these experiments.  We thank J. D. Edwards for the
computer program that generated the exact quantum mechanical wavefunctions.

\appendix
\section{Maslov indices}

\begin{figure}
    \begin{center}
        \epsfig{figure=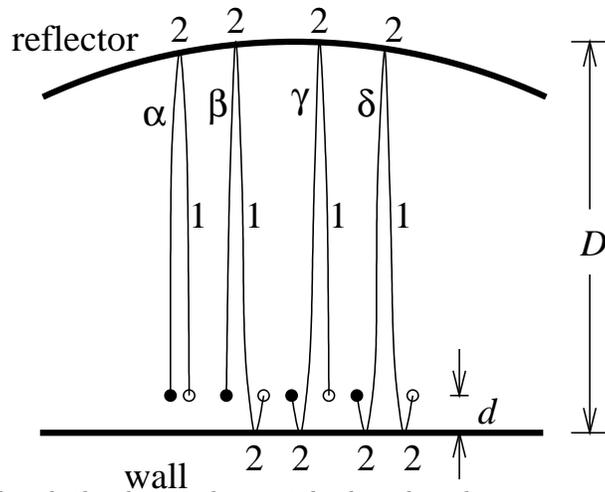,width=8cm}
        \caption[Four orbits]
        {Four orbits associated with the shortest horizontal orbit
          when the antenna is displaced from the wall.  The
          trajectories start from the filled circles and end on the
          open circles.  These circles are really the same point,
          namely the antenna position; they are displaced from each
          other in the figure so that the four distinct paths are
          visible.  The Maslov indices acquired on each segment of
          each orbit are indicated.  The antenna is a distance $d$
          away from the wall, and the horizontal distance from the
          wall to the reflector is $D$.}
        \label{4orbits}
    \end{center}
\end{figure}

In the experiment, the antenna is placed very close to, but not
exactly at, the wall.  This means there are in fact {\em four} orbits
associated with each single orbit in the billiard when the source is
placed exactly at the wall.  The four orbits associated with the
shortest horizontal orbit are shown in Fig.~\ref{4orbits}.
Each orbit in this family has a different total Maslov index and
length, summarized in Table~\ref{table:maslov}.
\begin{table}
    \begin{center}
        \begin{tabular}{|c|c|c|}
\hline
\multicolumn{3}{|c|}{\rule[-3mm]{0mm}{8mm}
  \bfseries Family of horizontal orbits}\\ 
  \hline 
  orbit & length & Maslov index \\
  \hline \hline
  $\alpha$ & $2(D-d)$ & 3 \\ \hline
  $\beta$ & $2D$ & 5 \\ \hline
  $\gamma$ & $2D$ & 5 \\ \hline
  $\delta$ & $2(D+d)$ & 7 \\ \hline
\end{tabular}
        \caption{Lengths and Maslov indices for the orbits 
          shown in Fig.~\ref{4orbits}.}
        \label{table:maslov}
    \end{center}
\end{table}
We can find the effect of grouping these four orbits into a single
orbit by computing the following sum,
\begin{equation}
    A e^{iS_{\rm eff} - i\pi \mu_{\rm eff}/2} = 
    \sum_{n=\alpha,\beta,\gamma,\delta} e^{iS_n - i\pi \mu_n/2},
\end{equation}
where the sum is over the four orbits as $d\to 0$.  The coefficient
$A$ will be found by doing the sum.  We have
\begin{eqnarray}
    \nonumber
    A e^{iS_{\rm eff} - i\pi \mu_{\rm eff}/2} &=& 
    e^{2ik(D-d) - 3\pi i/2} 
    +2e^{2ikD - 5\pi i/2}\\
    \nonumber
    &&+e^{2ik(D+d) -7\pi i/2}\\
    \nonumber
    &=&e^{2ikD - 5\pi i/2}\left(2-e^{2ikd}-e^{-2ikd}\right)\\
    \nonumber
    &=&e^{2ikD - 5\pi i/2}\left(2-2\cos(2kd)\right)\\
    &\approx& \left(2kd\right)^2e^{2ikD - 5\pi i/2}.
\end{eqnarray}
Thus we have that $A = (2kd)^2$, $\mu_{\rm eff} = 5$, and $S_{\rm eff}
= 2kD$, as expected.

\begin{figure}
    \begin{center}
        \centerline{\epsfig{figure=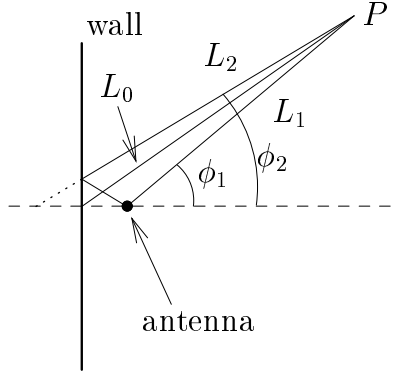,width=6cm}}
        \caption[Two trajectories]{Two trajectories arriving at the
          antenna from slightly different angles.  The antenna is a
          distance $d$ from the wall, as in the previous discussion.}
        \label{maslovangle}
    \end{center}
\end{figure}

A similar analysis on orbits coming to the antenna at an angle shows
that the coupling to the antenna varies as $\cos\phi$, as stated in
Eq.~(\ref{eq:sc_green_diff}).  Referring to Fig.~\ref{maslovangle},
a source of rays begins at the point $P$, and two of the rays find
their way to the antenna.  The direct path has length $L_1$ and Maslov
index $\mu_1 = 0$.  The second path first bounces off the wall before
arriving at the antenna, and has length $L_2$ and Maslov index $\mu_2
= 1$.  We want to combine these two paths into a single trajectory, by
doing a sum over the two trajectories as was done above, in order to
find the effective action and Maslov index for the trajectory.  We
further define the length $L_0$, which is the distance between $P$ and
the intersection of the wall with the axis of symmetry. The two
lengths $L_1$ and $L_2$ are given by
\begin{eqnarray}
    \nonumber
    L_{1,2} &=& \sqrt{\left(L_0 \cos\phi \mp d \right)^2 + \left(L_0 \sin\phi
    \right)^2}\\
    &\approx& L_0 \mp d\cos\phi,
\end{eqnarray}
where $\phi$ is the angle between the trajectory $L$ and the $x$-axis,
and $-$ and $+$ correspond to trajectories 1 and 2, respectively.  The
sum over orbits is then
\begin{eqnarray}
    \nonumber
    e^{iS_{\rm eff} - i\pi\mu_{\rm eff}/2} &=&
    e^{iS_1} + e^{iS_2 - i\pi} \\
    \nonumber
    &=& 2 \cos\left( k d\cos\phi - i \pi/2 \right) e^{i k L_0 - i\pi/2} \\
    &\approx& 2 kd \cos\phi e^{i k L_0 -i \pi/2}.
\end{eqnarray}
As expected, the effective action for the two paths is just $S_{\rm
  eff} = k L_0$, and the Maslov index is $\mu_{\rm eff} = 1$.  The
factor $\cos\phi$ appearing above is exactly the angular dependent
coupling coefficient appearing in Eq.~(\ref{eq:sc_green_diff}).  This
is just the simple angular dependence of a $p$-wave point source.

\section{Effect of a coarse probe}
\label{app:coarse-probe}

In this section we derive an effective potential describing the
perturbation to a TEM microwave resonance due to a coarse probe.  For
simplicity we consider the case of a closed cavity.  The electric
field ${\cal E}({\bf r},t) = \Real {\bf E}({\bf r}) e^{-i \omega t}$
and magnetic field ${\cal B}({\bf r},t) = \Real {\bf B}({\bf r}) e^{-i
  \omega t}$ of such a state can be expressed in terms of the quantum
analogue state, $\psi(x,y)$, as follows:
\begin{eqnarray}
    {\bf E}(x,y,z) & = & \frac{1}{h^{1/2}} \psi(x,y) \vhat{z} \\
    {\bf B}(x,y,z) & = & \frac{1}{h^{1/2}} \frac{1}{ik}
        \left[
            \frac{\partial \psi}{\partial y} \vhat{x} -
            \frac{\partial \psi}{\partial x} \vhat{y}
        \right]
        .
\end{eqnarray}
The height of the cavity in the $z$-direction is denoted by $h$.  (We
use the normalization conventions $\int \! dx \int \! dy \, \psi^2 =
\int \!  d^3\bvec{r} \, |\bvec{E}|^2 = \int \! d^3\bvec{r} \,
|\bvec{B}|^2 = 1$.)

The perturbation to the frequency of a microwave resonance by a
conducting probe is~\cite{maier54}
\begin{eqnarray}
    \frac{\omega^2 - \omega_0^2}{\omega^2} & = &
    \int_{\Delta V} \!\! d^3\bvec{r}
    \left( |\bvec{B}(\bvec{r})|^2 - |\bvec{E}(\bvec{r})|^2 \right) \\
    & = &
    \frac{1}{h} \int \!\! dx \! \int \!\! dy \, \delta h
    \left[
        \frac{1}{k^2} (\nabla \psi)^2
        - \psi^2
    \right]
    \label{eq:pert-integral}
    ,
\end{eqnarray}
where $\omega_0$ is the unperturbed frequency and $\delta h(x,y)$ is
the thickness of the probe.  The volume integral is over the volume
excluded by the probe; the second line follows because the fields have
no $z$-dependence.  From the chain rule and the Helmholtz equation,
\begin{eqnarray}
    \delta h (\nabla\psi)^2 & = &
    \nabla \cdot \left[
        \delta h \, \psi \nabla\psi
        - {\textstyle \frac{1}{2}} \nabla (\delta h) \psi^2
    \right] \nonumber \\
    & & {} + k^2 \, \delta h \, \psi^2
    + {\textstyle \frac{1}{2}} \nabla^2 (\delta h) \psi^2
    \label{eq:chainrule}
    .
\end{eqnarray}
Substitute Eq.~(\ref{eq:chainrule}) into Eq.~(\ref{eq:pert-integral}).
Since $\delta h$ is localized away from the lateral boundaries of the
cavity, the surface terms of Eq.~(\ref{eq:pert-integral}) vanish and
we are left with
\begin{equation}
    \frac{\omega^2 - \omega_0^2}{\omega^2} =
    \frac{1}{k^2} \int \!\! dx \! \int \!\! dy \,
    \frac{1}{2}\frac{\nabla^2(\delta h)}{h} \psi^2
    .
\end{equation}
Therefore, to first order in perturbation theory, the coarse probe can
be thought of as an effective perturbing potential of the form
\begin{equation}
    V_{\text{eff}}(x,y) \propto \nabla^2 \frac{\delta h(x,y)}{h}
    \label{eq:Veff}
    .
\end{equation}
Note that Eq.~(\ref{eq:Veff}) implies that the area integral of
$V_{\text{eff}}$ vanishes, which is a significant constraint on the
type of potential that can be modeled by a conducting probe in a
microwave cavity.  In particular, the perturbation of a 2DEG by an AFM
tip would not have this property.

In the case of our probe, $\delta h/h \approx f [ 1 - (\rho/\rho_0)^2
]$ where $\rho$ is the distance from the center of the probe, $\rho_0
= 40\text{ mm}$, and $f = 0.4$, so
\begin{equation}
    V_{\text{eff}} \approx
    -\frac{4 f}{\rho_0^2} + \frac{2 f}{\rho_0} \delta(\rho - \rho_0)
    ;
\end{equation}
in other words, the probe's effective potential is a flat well
surrounded by a repulsive ring at the probe's circumference.


\begin{thebibliography}{10}
\bibitem{footnote1} Note that the expression given in
    Ref.~\cite{james76} is the complex conjugate of the expression
    above, because the convention for the time dependence of the
    fields taken in Ref.~\cite{james76} is $e^{i\omega t}$, whereas we
    take the opposite convention, $e^{-i\omega t}$.
\bibitem{footnote2} The reason for the difference in the argument of
    the Fresnel integral is that in the half-line case the incident
    wave was plane ($l_1\to\infty$), whereas in the present case the
    incident wave is a cylindrical wave emanating from the QPC.
\end{thebibliography}
\end{document}